\documentclass[aps, amsfonts, amsmath, amssymb, showpacs,showkeys,onecolumn,preprint,preprintnumbers]{revtex4}
\usepackage{latexsym}
\usepackage{amssymb}
\usepackage{amsfonts}
\usepackage{graphicx,color}
\usepackage{graphicx}
\usepackage{slashbox}
\usepackage{mdframed}
\usepackage{ifthen}
\usepackage{amsthm}

\usepackage{amsmath}
\numberwithin{equation}{section}

\makeatletter
\renewcommand{\p@subsection}{}
\renewcommand{\p@subsubsection}{}
\makeatother

\usepackage{hyperref}\providecommand\USINGhyperrefNOW{true}

\newtheorem{theorem}{Theorem}[section]

\begin{document}

\title{Possible Consistent Extra Time Dimensions in the Early Universe}
\author{Patrick L. Nash}
\email{Patrick299Nash@gmail.com}
\affiliation{
475 Redwood Street Unit 801, San Diego, CA 92103-5864
}

\date{\today}

\begin{abstract}
Gravity cannot be quantized unless the quantized  theory is cast on a manifold whose
concomitant
number of physical space dimensions
and number of physical time dimensions correspond to physical reality, and not simply to the perception of reality.
At present, the accepted number of physical time dimensions is dictated more by folklore than by science.
True, there are theorems that restrict the number of possible time dimensions to one, but these only apply to
rather special scenarios. In this paper we discuss a model of the early universe
in which the number of physical time dimensions is four, and formulate  Theorem[\ref{tj}],
which underlies an explanation of why the extra time dimensions
do not source unphysical effects.

In this paper we describe a  new model of gravitational inflation
that is driven by dark energy
and ``mediated" by a real massless scalar inflaton field $\varphi$
whose potential is identically equal to zero.
The coupled Einstein gravitational and inflaton field equations are
formulated on an eight-dimensional spacetime manifold
of \textbf{four space} dimensions and \textbf{four  time} dimensions.
We find explicit solutions to these field equations
that exhibit
temporal exponential
\textbf{deflation of three of the four time dimensions},
and then study the dynamics of a massive complex scalar field $\psi$ that
propagates on the background ground state Einstein gravitational field
to determine whether its quantum fluctuations  $\delta \psi$ are stable or unstable.
We compute explicit approximate solutions to the  $\delta \psi$  field equations
that are \textbf{stable}, meaning that the quantum fluctuations  $\delta \psi$ of the field  $\psi$
do not grow exponentially with time.
\textbf{Instabilities}
driven by the momenta associated to the three extra time dimensions do not appear in the
physical solutions of the field equations of this model.
\end{abstract}

\maketitle

\section{Introduction}
\label{sec:introduction}
Recent Planck 2013 data analysis \cite{Ade2013} is in remarkable accord with
a flat $\Lambda$CDM model with inflation, based
upon a spatially flat, expanding universe whose dynamics are
governed by General Relativity and sourced by cold dark matter,
a cosmological constant $\Lambda$,
and a slow-roll scalar inflaton field \cite{Guth:1980zm,Linde:1981mu,Albrecht:1982wi}.
The main predictions of  inflationary  cosmology  are also
consistent \cite{Dunkley:2008ie,Komatsu:2008hk,0067-0049-192-2-18,Kinney:2008wy}
with other recent observational data from important
experiments such as WMAP
\cite{Hinshaw.1212.5226v2}
and the Sloan Digital Sky
Survey \cite{PhysRevD.69.103501,0004-637X-633-2-560,PhysRevD.74.123507}, to name only two.
Also, assuming that
the large value of the tensor-to-scalar ratio in the cosmic microwave background radiation
reported by the BICEP2 collaboration \cite{Ade2014xna} is correct,
then the energy scale of inflation is approximately $2\times10^{16}$ GeV,
which is only
roughly two orders of magnitude below the Planck scale of $\frac{1.22\times 10^{19}}{\sqrt{8 \pi }}$ GeV.

However, parameterized $\Lambda\-$CDM inflationary  cosmology may not represent fundamental physics.
Here we discuss a radical new model of inflation/deflation that overcomes the
well known issues
of
[1] the precise definition of the inflaton potential,
[2] the precise definition of the inflaton mass
(many models of inflation omit the Higgs field that presumably generates the inflaton mass,
or do not explain why the Higgs field develops a non-zero vacuum expectation value at such a high energy) and
[3] re-heating.
This model is based on the idea that our universe has as many time dimensions
as space dimensions.

In this paper we describe a  new model of gravitational inflation
that is driven by dark energy
and ``mediated" by a real massless scalar inflaton field $\varphi$
whose potential is identically equal to zero.
The coupled Einstein gravitational and inflaton field equations are
formulated on an eight-dimensional spacetime manifold
of \textbf{four space} dimensions and \textbf{four  time} dimensions.
For the case of a diagonal metric, two periodic solution classes
(``ground" and ``excited" classes)
for the coupled Einstein field equations are obtained that exhibit
temporal exponential
\textbf{deflation of three of the four time dimensions}
and temporal exponential inflation
of three of the four space dimensions.
Moreover, for the ground state solution class \textbf{the universe does not cool during inflation}.
Reheating \cite{PhysRevD.56.3258} is a non-issue.
Moreover, we show that
the extra time dimensions do not generally induce  exponentially rapid growth of fluctuations in quantum fields
that propagate on the ground state solution class.

Comoving coordinates for the two non-inflating/deflating dimensions are chosen to be
$(x^4  , x^8  )$.
The $x^4$ coordinate corresponds to our universe's observed physical time dimension, while
the $x^8$ coordinate corresponds to a new spatial dimension that is assumed to be compact
with constant unit radius $R_8\,\equiv\,1$, i.e., the radius equals the Planck length.
The   value of $R_8$ can be adjusted to fit experimental data if need be.
Dark energy is
realized in terms of a cosmological constant $\Lambda$,
whose values are determined to be quantized
in terms of the constant unit radius
of the compact spatial $x^8$ dimension.
It is also interesting to note that
although the inflaton potential is zero,
an effective inflaton potential for the ground state
is computed to be
$V_{\textrm{eff}} = \frac{5}{4!} \, \Lambda \, {\cot{(\frac{1}{2}x^8 \, \sqrt{2 \Lambda}) }}^2$.

In this model, after ``inflation" the observable physical macroscopic world appears to a classical observer to
be a homogeneous, isotropic universe with three space dimensions and one time dimension.
Also, under the assumption that an ``arrow of time" exists for each timelike dimension,
unphysical closed timelike curves are not observed in this model.
We also prove that a well known so-called ``single time" theorem \textbf{does not apply}
to our model.

\section{Arguments against plural time dimensions}
Everyone is familiar with arguments against the existence of extra time dimensions.
While there are many aspects to this issue, we address two arguments against the existence of extra time dimensions
which stand out most clearly\cite{linde1990particle}:

\begin{enumerate}
\label{eissues}
\item {A spacetime with extra time dimensions may not carry a spin structure $\Leftrightarrow$ spinors cannot be defined on the spacetime}\label{sss1};
\item {Momenta corresponding to the extra time dimensions induce exponentially rapid growth of quantum fluctuations of the field; the universe is unstable.
This instability is associated with the very largest momenta (shortest wavelengths).
}\label{sss2}
\end{enumerate}
Issue [\ref{sss1}] is not applicable here; it is  known that our model spacetime carries a spin structure \cite{Nash2010}.
The issues raised in [\ref{sss2}]
may be investigated by studying the field equation for the propagation of
a massive complex scalar field $\psi$ on the background gravitational field:
\begin{equation}
\label{escact}
0=\frac{1}{\sqrt{det(g_{\alpha\,\beta})}} \; \partial_{\mu}
\left[ \;
\sqrt{det(g_{\alpha\,\beta})} \; g^{\mu\,\nu}\;\partial_{\nu}\psi\; \right]
- \;\left(\Lambda\, m^2\,+\,\zeta\,R\right)\;\psi
,
\end{equation}
which is derivable from the Lagrangian
\begin{equation}
\label{scact}
L_{\psi}=\sqrt{det(g_{\alpha\,\beta})}\left[-\; g^{\mu\,\nu}\partial_{\mu}\psi^{*}\;\partial_{\nu}\psi\;
-\psi^{*}\;\psi\;\left(\Lambda\, m^2\,+\,\zeta\,R\right) \; \right]
.
\end{equation}
Here $R$ is the Ricci scalar, and $m$ and $\zeta$ are real input parameters;
the factors of $\Lambda$ are included for convenience.
The quantum fluctuations  $\delta \psi$ of $\psi$ satisfy a field equation similar to Eq.[\ref{escact}],
but generally with a different rest mass parameter.

It will be seen that  Eq.[\ref{escact}] is not separable.
Because Eq.[\ref{escact}] is not separable, there will not be a simple
dispersion relation, \emph{per se}, relating a physical-frequency ($\omega \leftrightarrow - \dot{\imath}\frac{\partial}{\partial \, x^4}\ln{\psi}$) to
the momentum wave vectors $(\vec{k}, \vec{w})$ that are defined two paragraphs below.
In fact we find decaying quasi-normal mode approximate solutions to   Eq.[\ref{escact}].
Modes of this class  are well known in black hole physics \cite{Dyatlov2011, lrr-1999-2,doi:10.1142/S021773230401240X}:
the authors of
Ref.[\onlinecite{lrr-1999-2}]  discuss
the theory of quasi-normal modes of compact objects,
including perturbations of (Schwarzschild, Reissner-Nordstr{\"o}m, Kerr
and Kerr-Newman) black holes;
Ref.[\onlinecite{Dyatlov2011}]
provides a rigorous definition of quasi-normal modes for a rotating black hole;
the author proves that the local energy of linear waves in certain black hole backgrounds decay exponentially once
orthogonality to the zero resonance is imposed.
In the present early universe model, decaying quasi-normal type modes  appear as approximate solutions to   Eq.[\ref{escact}];
however numerical experiments indicate that the contribution to the complex
frequency that is responsible for decay gets smaller as the order of the approximation increases
(at least for a finite percentage of the modes).
The consequence of this for the unitarity of the quantum field theory is not presently known.

We shall discuss two types of solution to  Eq.[\ref{escact}], the first with nonzero coupling to the Ricci scalar
and the second with $\zeta = 0$.
It should be emphasized that the expected Yukawa interaction coupling of the massive complex scalar field $\psi$
with the massless real scalar inflaton  field $\varphi$ is not included here.
We do not wish to confuse the production/annhilation  of  $\psi$ ``particles"
due to its interaction with the inflaton field
with instabilities in the  $\psi$ field that are sourced by the momenta associated to the extra time dimensions.
Also, it is beyond the scope of the present paper to consider the
one-loop effective potentials and concomitant renormalizations
of both $\varphi$ and $\psi$. These are  important questions, but outside the scope of this work.

In Sections \ref{sec:C0} and \ref{sec:C1} we  seek ``plane wave" solutions to  Eq.[\ref{escact}]
in terms of comoving wave vectors and coordinates of the form

\begin{eqnarray}
\label{spsi}
\psi
&=&
\Psi(x^4, x^8) \,  {e}^{\dot{\imath}\left(\vec{k} \cdot \vec{r} - \vec{w} \cdot \vec{R} \right)},\;\textrm{ where }
\nonumber \\
\vec{r} &=& \left(x^1, x^2, x^3\right)^{T}\;\textrm{ and }
\vec{k} = \left(k_1, k_2, k_3\right)^{T}\;;
\nonumber \\
\vec{R} &=& \left(x^5, x^6, x^7\right)^{T}\;\textrm{ and }
\vec{w} = \left(k_5, k_6, k_7\right)^{T}
.
\end{eqnarray}
When discussing these solutions the following shorthand is employed:
\begin{eqnarray}
\label{kspsi}
\Lambda\, k^2 \, &=&\vec{k} \cdot  \vec{k} = \,{k_1}^{2}+{k_2}^{2}+{k_3}^{2}\;;
\nonumber \\
\Lambda\, w^2 \,&=&\vec{w} \cdot  \vec{w} = \, {k_5}^{2}+{k_6}^{2}+{k_7}^{2}
.
\end{eqnarray}

To summarize, in this paper we find explicit solutions of the coupled Einstein-inflaton field equations,
which are then substituted into  Eq.[\ref{escact}];
solutions to the latter equation are then determined for which the effects of multiple time dimensions
are stable, meaning that quantum fluctuations  $\delta \psi$ of the $\psi$ field do not grow exponentially with time.
Exponentially rapid growth of quantum fluctuations  $\delta \psi$ of the $\psi$ field  cause perturbation theory to break down when
the norm of the term(s) involving $\delta \psi$ become of order one.

\subsection{Single-Time Theorem (STT)}

The following ``Single-Time" Theorem \cite{:/content/aip/journal/jmp/40/2/10.1063/1.532695} has  been asserted, incorrectly, to argue that
extra time dimensions  cannot exist in viable physical  models of the  universe.
To demonstrate that this theorem does not apply to the model in this paper, we need only quote the statement
of the STT and highlight two points in its proof
that clearly are not satisfied in our model.
We quote the author of the STT:

``Theorem 2 (Single-Time Theorem) \cite{:/content/aip/journal/jmp/40/2/10.1063/1.532695}. Any BLACK HOLE solution with $k > 0$ contains precisely one coordinate $t$ such
that $g_{tt} = 0$ at the horizon."

``The starting point is ... the model action for D-dimensional gravity with several scalar dilatonic fields $\phi_a$ and antisymmetric
$n_s$-forms $F_s$ ...
in a (pseudo-)Riemannian manifold $M= R_u \times M_0 \times \ldots \times M_n$ with the metric
$ds^2 = g_{M N} dz^M dz^N = w e^{2 \alpha(u)} du^2 + \sum_{i=0}^n \,e^{2 \beta^i(u)}ds_i^{2} , \; w = \pm 1$,
where $u$ is a selected coordinate ranging in $R_u \subseteq R; g^i = ds_i^{2}$
are metrics on $d_i$ -dimensional factor spaces $M_i$ of
arbitrary signatures
$\varepsilon_i = \textrm{sign  } g^i ; |g| = | det g_{M N}|$
and similarly for subspaces; ...
$M_i , i > 0$ are assumed to be Ricci-flat, while $M_0$ is allowed to be a space of constant curvature $K_0 = 0, \pm1$."

Two key steps in the proof of the STT, which have no counterpoint in our model, are the facts that the author of the STT :
$[1]$ employs the equations of motion of a test particle moving in the presence of a black hole (see STT Eq.[29]);
and
$[2]$  employs the black hole potential (see STT Eq.[30])
in order
to derive several equalities, from which he then
finally concludes that more than one time dimension is not possible for his model.

However, in our present model the very early universe is modeled as \emph{homogeneous} and isotropic.
There are no black holes and no black hole potentials, so the STT clearly does not apply.

\section{Notation and conventions}
This model is cast on  $\mathbb{X}_{4,4}$, which is an eight-dimensional pseudo-Riemannian manifold
that  is a spacetime of four space dimensions, with local comoving spatial coordinates $(x^1, x^2, x^3, x^8)$,
and four  time dimensions,  with local comoving temporal coordinates  $(x^4, x^5, x^6, x^7)$
(employing the usual component notation in local charts).
Greek indices run from 1 to 8.
Each of the  $\, x^{\alpha}\,$ has units (dimensions) of length.
It is assumed that $-\infty < \, x^{\alpha}\, <\infty$
for $\alpha = 1, 2, \ldots, 7$.

The eight-dimensional pseudo-Riemannian manifold $\mathbb{X}_{4,4}$ may be decomposed as
$\mathbb{X}_{4,4} = \mathbb{X}_{3,4} \times S^{1}_{(8)}$, where $S^{1}_{(8)}$ is  a one dimensional compact space that is
homeomorphic to the unit circle $S^1$.
We need at least two charts to cover  $S^{1}_{(8)}$, which are chosen to be
$x^8 : S^{1}_{(8)}\,\rightarrow \mathbb{R} $ ,  $\;\;-\pi \, R_8< \, x^{8}\, <\pi\,R_8 \;$
and $x'^8 : S^{1}_{(8)}\,\rightarrow \mathbb{R}$,  $0\,< \, x'^{8}\, <\,2\,\pi\,R_8$,
where $R_8\,\equiv\,1$ is the Planck length. $R_8$ is introduced
to give $x^8$ the correct dimension.  $R_8$ is not associated to a dilatonic degree of freedom
since it is constant, $R_8\,\equiv\,1$.

Let $U$ be a subset of $\mathbb{X}_{4,4}$, $p \in U$ and $\phi : U \rightarrow \mathbb{R}^8$ be a local chart.
$\frac{\partial\phantom{\,x}}{\partial\,x^{\alpha}}$ denotes the natural basis of $ T_p(\mathbb{X}_{4,4} )$ associated with the coordinates
$x^{\alpha} =\left[\phi(p)\right]^{\alpha}$,
where $\left[.\right]^{\alpha}$ denotes the canonical projection of the ${\alpha}^{\textrm{th}}$ component
of $x \in \mathbb{R}^8$.
In the context of this paper a phrase such as ``$x^6$ is a timelike dimension" means
that the timelike vector $\frac{\partial\phantom{\,x}}{\partial\,x^6}$ is tangent to the lines of constant spatial coordinates
and constant $(x^4, x^5, x^7)$, i.e. the curves
of  $\mathbb{X}_{4,4}$ defined by
$(x^1 = K_1, x^2 = K_2, x^3 = K_3, x^4 = K_4, x^5 = K_5, x^7 = K_7,  x^8 = K_8)$, where
$K_1,  K_2,  K_3,  K_4, K_5,  K_6,   K_8$ are seven constants.

In this paper we obtain solutions to the coupled Einstein-inflaton field equations that exhibit
temporal exponential deflation of three of the four time dimensions.
More precisely, we record solutions that possess exponentially decreasing scale factors for
three of the four timelike dimensions, where decreasing means  decreasing with
respect to the remaining fourth time coordinate,
which is chosen to be $x^4$.
In effect we arbitrarily label the temporal direction/dimension that does not deflate (or inflate, for that matter)
the $x^4$- axis. The three remaining time dimensions
are labeled with $(x^5, x^6, x^7)$
and they undergo deflation.
We assume that unknown quantum gravity effects
sort things out shortly after the Universe is created so that the three mutually orthogonal
deflating timelike directions and the
remaining fourth mutually orthogonal
non-inflating/non-deflating  timelike direction (i.e., the temporal dimension whose scale factor is a constant)
are the same for all
values of $(x^1, x^2, x^3, x^8)$.
This is an unsatisfying assumption, but a necessary hypothesis given our current knowledge of quantum gravity.

Let $\mathbf{g}$ denote the pseudo-Riemannian metric tensor on $\mathbb{X}_{4,4}$.
The signature of the  metric $\mathbf{g}$ is
$(4,4) \leftrightarrow (+ + + - - - - +)$.
The   covariant derivative with respect to the
symmetric connection  associated to the metric $\mathbf{g}$ is denoted by a vertical double-bar.
We employ the Landau-Lifshitz spacelike sign conventions \cite{Misner1973}.
$\mathbf{g} \leftrightarrow g_{ \alpha \beta  }= g_{ \alpha \beta  }(x^{\mu})$
is assumed to carry the Newton-Einstein gravitational degrees of freedom.
It is moreover assumed that the ordinary Einstein field equations (on $\mathbb{X}_{4,4}$)
\begin{equation}
\label{EFQ}
G_{\alpha \beta}  \;+\,g_{ \alpha \beta}\,\Lambda\, = \; 8 \;  \pi \;  \mathbb{G} \; T_{\alpha \beta}
\end{equation}
are satisfied.
Here
$G_{\alpha \beta}$ denotes the Einstein tensor, $ \Lambda$ is the cosmological constant,
$\mathbb{G}$ denotes  the Newtonian gravitational constant, and the
reduced Planck mass is $M_{Pl} = \left[ \; 8 \pi \, \mathbb{G} \;  \right]^{-1/2}$.
Natural units $c = 1 = \hbar\,=\, \; 8 \pi \, \mathbb{G} $ are used throughout.
Lastly, if $f = {f(x^{4},x^{8})}$ then
\begin{eqnarray}
\label{fxxx0}
\begin{array}{ccc}
f^{(1,0)}=\frac{\partial }{\partial \, x^{4}} \, f(x^{4},x^{8}),  &
f^{(0,1)}=\frac{\partial }{\partial \, x^{8}} \, f(x^{4},x^{8}),  &
f^{(1,1)}=\frac{\partial^{2} }{\partial \, x^{4}\, \partial \, x^{8}} \, f(x^{4},x^{8}), \\
f^{(2,0)}=\frac{\partial^{2} }{\partial \, {x^{4}}^{2}} \, f(x^{4},x^{8}), &
f^{(0,2)}=\frac{\partial^{2} }{\partial \, {x^{8}}^{2}} \, f(x^{4},x^{8}), &
\textrm{ etc.}
\end{array}
\end{eqnarray}

\section{Einstein field equations}

\subsection{Preview}
We seek solutions to the  Einstein field equations Eq.[\ref{EFQ}]
for which the observable physical macroscopic early universe appears to a classical observer, after inflation ends, to
be a homogeneous, isotropic universe with three space dimensions and one time dimension.
Accordingly the three space dimensions $(x^1, x^2, x^3)$ are assumed to possess equal scale factors, denoted $a = a(x^4, x^8)$,
and   the three extra time dimensions $(x^5, x^6, x^7)$ are assumed to possess equal scale factors, denoted $b = b(x^4, x^8)$.

We find two periodic (in $x^8$) solution classes
$( E_0 , E_1)$ to the  Einstein field equations Eq.[\ref{EFQ}] on $\mathbb{X}_{4,4}$
that exhibit inflation/deflation and
describe a  universe that is spatially flat throughout the inflation era.
Moreover it is found
that $\Lambda$ is quantized in terms of  $R_8$.
For $( E_0 , E_1)$, during ``inflation",
the scale factor $a = a(x^4, x^8)$ for the three   space dimensions $(x^1, x^2, x^3)$
exponentially inflates as a function of $x^4$,
and
the scale factor  $b = b(x^4, x^8)$ for the three extra time dimensions $(x^5, x^6, x^7)$
exponentially deflates as a function of $x^4$;
moreover, the scale factors for the   $x^4$ and $x^8$ dimensions are constants equal to one.

We define the
cosmological \emph{ relative expansion rate} $\ell$ as
\begin{equation}
\label{LR}
\ell =  \frac{1}{2} \frac{\partial}{\partial x^4} \ln{\left( \frac{b}{a} \right)}
.
\end{equation}
In Section \ref{sec:C0} it is shown
that $\ell$ is independent of time and equal to
$\ell_0 = -  \sqrt{ \frac{1}{18} \Lambda }$
for solutions in the first class $E_0$;
this solution corresponds to pure exponential time inflation of
the scale factor  $a(x^4, x^8)$
and pure exponential time deflation of
the scale factor  $b(x^4, x^8)$.
For this case $|\ell_0|$ coincides with the Hubble parameter $H$.

For the class  $E_0$ solution we find that
$\Lambda $ is ``quantized" in terms of the
compact
spatial $x^8$ unit radius   $R_8\,\equiv\,1$, and given by
\begin{equation}\label{LC}
\Lambda_0\,=\, \frac{1}{2} \, \frac{1}{{R_{8}}^2},
\end{equation}
and therefore is one half the square of the Planck mass.

Therefore $\ell_0$ verifies
\begin{equation}
R_8 \, |\ell_0| = \frac{1}{6}
,
\end{equation}
The  two relationships are reminiscent of  semiclassical quantization relationships.

For solutions in the second class $E_1$, which describes a  transition from pure exponential ``inflation,"
$\ell =\ell_1$ is generally time dependent.
In this case the ``quantized" value of
$\Lambda $ is
\begin{equation}\label{LC2}
\Lambda_1= \frac{6}{5}\Lambda_0.
\end{equation}
We refer to $\Lambda_0$ as the ground state value of the cosmological constant and
$\Lambda_1$ as its first excited state.

Mathematical universes that possess positive relative expansion rate $\ell > 0$ for $x^4 > 0$ exist
and are reported below.
However they are analogous to collapsing universe solutions in the usual Friedmann cosmological model,
or to universes with multiple macroscopic time dimensions,
and hence are not directly relevant to the physical processes discussed in this paper.

\subsection{Line element}
The line element for inflation/deflation is assumed to be given by
\begin{eqnarray}
\label{dsds}
\left\{
ds
\right\}^2
&=&
\left\{
a( x^4, x^8 )
\right\}^2
\left[
{\left(dx^1\right)}^2 + {\left(dx^2\right)}^2 +{\left(dx^3\right)}^2
\right]
-
{\left( dx^4\right)}^2
\nonumber \\
&-&
\left\{
b( x^4, x^8 )
\right\}^2
\left[
{\left(dx^5\right)}^2 + {\left(dx^6\right)}^2 + {\left(dx^7\right)}^2
\right]
+ {\left( dx^8\right)}^2
\nonumber \\
&=&
{a}^{2}
\left[
{\left(dx^1\right)}^2 + {\left(dx^2\right)}^2 +{\left(dx^3\right)}^2
\right]
-
{b}^{2}
\left[
{\left(dx^5\right)}^2 + {\left(dx^6\right)}^2 + {\left(dx^7\right)}^2
\right]
\nonumber \\
&-&
{\left( dx^4\right)}^2
+ {\left( dx^8\right)}^2
\;\;;
\end{eqnarray}
where $a = a(x^{4},x^{8})$ and $b = b(x^{4},x^{8})$ carry the metric  degrees of freedom in this model.
The real massless scalar inflaton field is
$\varphi = \varphi(x^{4},x^{8})$.
The action for the metric and inflaton degrees of freedom is assumed to be given by
\begin{equation}
\label{act}
S = \int \;
{
\left[
\frac{1}{16\,\pi\,\mathbb{G}} \left(R- 2 \Lambda \right)
\;-\;
\frac{1}{2}g^{\mu\,\nu}\partial_{\mu}\varphi\;\partial_{\nu}\varphi\;
\;-\;\xi\,R\,\varphi^2
\;+\;
\mathfrak{L}_{\textrm{SM}}
\right]
}^2 \;
\sqrt{det(g_{\alpha\,\beta})}
\; d^{8}x
.
\end{equation}
Here $ \Lambda$ is the cosmological constant.
The inflaton potential is zero; its action is purely kinematic, although
$\varphi ^{(0,1)}(x^4,x^8)^2$ may  be regarded as contributing to an effective inflaton potential.
For the remainder of the paper we assume minimal coupling of the inflaton with gravity and set $\xi = 0$.
We also neglect the standard model degrees of freedom and set $\mathfrak{L}_{\textrm{SM}}$ = 0.

\subsection{Canonical stress-energy tensor}
The canonical stress-energy tensor for the real massless scalar inflaton field is
$T_{{\mu}\, \nu}=
-\frac{2}{\sqrt{\textrm{det}(g_{\alpha\beta})}}
\frac{\partial}{\partial\,g^{{\mu}\, \nu}}L_{\varphi}
$,
$
L_{\varphi}=\sqrt{det(g_{\alpha\,\beta})}\left[-\;\frac{1}{2}g^{\mu\,\nu}\partial_{\mu}\varphi\;\partial_{\nu}\varphi\; \right]
$.
The distinct  components are

\begin{eqnarray}
\label{Tx33}
T_{3\,3}
&=&
\frac{1}{2} \mathbf{a}(x^4,x^8)^2 \left[\,\varphi ^{(1,0)}(x^4,x^8)^2-\varphi ^{(0,1)}(x^4,x^8)^2\right],
\end{eqnarray}

\begin{eqnarray}
\label{Tx44}
T_{4\,4}
&=&
\frac{1}{2} \left[\,\varphi ^{(1,0)}(x^4,x^8)^2+\varphi ^{(0,1)}(x^4,x^8)^2\right],
\end{eqnarray}

(which clearly satisfies the “weak energy condition”),

\begin{eqnarray}
\label{Tx55}
T_{5\,5}
&=&
\frac{1}{2} \mathbf{b}(x^4,x^8)^2 \left(\,\varphi ^{(0,1)}(x^4,x^8)^2-\varphi ^{(1,0)}(x^4,x^8)^2\right),
\end{eqnarray}

\begin{eqnarray}
\label{Tx88}
T_{8\,8}
&=&
\, \frac{1}{2} \varphi ^{(0,1)}(x^4,x^8)^2+\frac{1}{2} \varphi ^{(1,0)}(x^4,x^8)^2,
\end{eqnarray}
and

\begin{equation}
\label{Tx48}
T_{4\,8}
=
T_{8\,4}
=
\varphi ^{(0,1)}(x^4,x^8) \varphi ^{(1,0)}(x^4,x^8),
\end{equation}
which is non-zero, in general.
Due to the $(x^4,x^8)$ dependence of the metric the $(4,8)$ and  $(8,4)$
components  of the Einstein tensor $G_{4\,8}=G_{8\,4}$  are also, in general,
non-zero.

Let $\rho = \rho(x^4,x^8)$ denote the effective energy density
of the (classical) inflaton field
and $p = p(x^4,x^8)$ denote the effective pressure.
$\rho $ and  $p$ may be associated to  $T_{{\mu}\, \nu}$ according to
\begin{equation}
\label{Tx48eff}
T^{{\mu}\, \nu}
=
\left[\sum_{j=4}^{j=7}\;\;u_{(j)}^{\mu}\;u_{(j)}^{\nu}\right]\,\left(\rho + p\right)\;+\,p\,g^{{\mu}\, \nu}
\;+
\left[\,\delta^{\mu}_{4}\delta^{\nu}_{8}+\delta^{\mu}_{8}\delta^{\nu}_{4}\right]T^{4\, 8}
\end{equation}
where
$u_{(j)}^{\mu} = \delta^{\mu}_{j}\;j = 4, 5, 6, 7$ in this comoving frame.
This implies that
\begin{eqnarray}
\label{Tx44x}
T_{4\,4} &=& \rho =
\frac{1}{2} \left[\,\varphi ^{(1,0)}(x^4,x^8)^2+\varphi ^{(0,1)}(x^4,x^8)^2\right]
\nonumber \\
T_{3\,3}&=& g_{3\,3} \,p =
\frac{1}{2} \mathbf{a}(x^4,x^8)^2 \left[\,\varphi ^{(1,0)}(x^4,x^8)^2-\varphi ^{(0,1)}(x^4,x^8)^2\right].
\end{eqnarray}
Therefore
\begin{equation}
\label{w}
w =
\frac{p}{\rho} =
\frac{T_{3\,3}}{g_{3\,3}\,T_{4\,4}} =
\frac{\varphi ^{(1,0)}(x^4,x^8)^2-\varphi ^{(0,1)}(x^4,x^8)^2}{\varphi ^{(1,0)}(x^4,x^8)^2+\varphi ^{(0,1)}(x^4,x^8)^2}.
\end{equation}
When
$\varphi ^{(1,0)}(x^4,x^8) = 0$,
as is the case for the Class $E_0$ solution of Subsection[\ref{sec:C0}],
then
\begin{equation}
\label{w1}
w =-1 .
\end{equation}
In this case the effective inflaton potential
may be defined as
\begin{equation}
\label{Veff}
V_{\textrm{eff}} = \frac{1}{2}\varphi ^{(0,1)}(x^4,x^8)^2.
\end{equation}
Therefore for the Class $E_0$ solution of Section[\ref{sec:C0}],
\begin{equation}
\label{Veff0}
V_{\textrm{eff}} = \frac{5}{4!} \, \Lambda \, {\cot{(\frac{1}{2}x^8 \, \sqrt{2 \Lambda}) }}^2.
\end{equation}

\subsection{Field Equations}
\begin{equation}
G_{\mu \, \nu}
\,+\, \Lambda\,g_{ {\mu} \, \nu}
\; = \; 8 \;  \pi \;  \mathbb{G} \; T_{ {\mu} \, \nu}
\end{equation}
The distinct field equation components may be written as

\begin{equation}
\label{x48}
G_{4\,8}
=
G_{8\,4}
=
-\frac{3 a^{(1,1)}{}}{a{}}-\frac{3 b^{(1,1)}{}}{b{}}
=
8 \pi  \mathbb{G} \; \varphi ^{(0,1)}{} \varphi ^{(1,0)}{}
\end{equation}

\begin{eqnarray}
\label{x33}
G_{3\,3}
&=&
\frac{1}{b{}^2}
\left[
3 a{} b{} \left(2 a^{(0,1)}{} b^{(0,1)}{}-2 a^{(1,0)}{} b^{(1,0)}{}+a{} \left(b^{(0,2)}{}-b^{(2,0)}{}\right)\right)
\right.
\nonumber \\
&+&
\left.
\left(a^{(0,1)}{}^2-a^{(1,0)}{}^2+2 a{} \left(a^{(0,2)}{}-a^{(2,0)}{}\right)\right) b{}^2+3 a{}^2 \left(b^{(0,1)}{}^2-b^{(1,0)}{}^2\right)
\right]
\nonumber \\
&=&
4 \pi  \mathbb{G} a{}^2 \left( \, -\varphi ^{(0,1)}{}^2+\varphi ^{(1,0)}{}^2\right)
\end{eqnarray}

\begin{eqnarray}
\label{x44}
G_{4\,4}
&=&
-\frac{3}{a{}^2 b{}^2}
\left[
b{}^2
\left(a^{(0,1)}{}^2-a^{(1,0)}{}^2+a{} a^{(0,2)}{}\right)
+
a{} b{} \left(3 a^{(0,1)}{} b^{(0,1)}{}-3 a^{(1,0)}{} b^{(1,0)}{}+a{} b^{(0,2)}{}\right)
\right.
\nonumber \\
&+&
\left.
a{}^2 \left(b^{(0,1)}{}^2-b^{(1,0)}{}^2\right)
\right]
\nonumber \\
&=&
4 \pi  \mathbb{G} \left( \,\varphi ^{(0,1)}{}^2+\varphi ^{(1,0)}{}^2\right)
\end{eqnarray}

\begin{eqnarray}
\label{x55}
G_{5\,5}
&=&
\frac{1}{a{}^2}
\left[
2 a{} b{} \left(-3 a^{(0,1)}{} b^{(0,1)}{}+3 a^{(1,0)}{} b^{(1,0)}{}+a{} \left(b^{(2,0)}{}-b^{(0,2)}{}\right)\right)
\right.
\nonumber \\
&+&
\left.
3 \left(-a^{(0,1)}{}^2+a^{(1,0)}{}^2+a{} \left(a^{(2,0)}{}-a^{(0,2)}{}\right)\right) b{}^2+a{}^2 \left(b^{(1,0)}{}^2-b^{(0,1)}{}^2\right)
\right]
\nonumber \\
&=&
4 \pi  \mathbb{G} b{}^2 \left(\,\varphi ^{(0,1)}{}^2-\varphi ^{(1,0)}{}^2 \right)
\end{eqnarray}

\begin{eqnarray}
\label{x88}
G_{8\,8}
&=&
-\frac{3}{a{}^2 b{}^2}
\left[
a{}^2 \left(b^{(1,0)}{}^2-b^{(0,1)}{}^2\right)
+
\left(-a^{(0,1)}{}^2+a^{(1,0)}{}^2+a{} a^{(2,0)}{}\right) b{}^2
\right.
\nonumber \\
&+&
\left.
a{} b{} \left(-3 a^{(0,1)}{} b^{(0,1)}{}+3 a^{(1,0)}{} b^{(1,0)}{}+a{} b^{(2,0)}{}\right)
\right]
\nonumber \\
&=&
4 \pi  \mathbb{G} \left( \, \varphi ^{(0,1)}{}^2+\varphi ^{(1,0)}{}^2\right)
\end{eqnarray}

The components of $T^{\mu}_{||{\mu}\; \alpha}$ that are not identically zero must satisfy
\begin{eqnarray}
\label{xxxt}
a\,b\;T^{\mu}_{||{\mu}\; 4}
&=&
0\,=
3 a{} \left(-b^{(1,0)}{} \left( \varphi ^{(1,0)}{}^2 \right)+b^{(0,1)}{} \varphi ^{(0,1)}{} \varphi ^{(1,0)}{}+b^{(0,1)}{}  \right)
\nonumber \\
&+&
b{} \left(-3 a^{(1,0)}{} \left( \varphi ^{(1,0)}{}^2\right)+3 a^{(0,1)}{} \varphi ^{(0,1)}{} \varphi ^{(1,0)}{}
\right.
\nonumber \\
&+&
\left.
a{} \left( \varphi ^{(1,0)}{} \left(\, \varphi ^{(0,2)}{}-\varphi ^{(2,0)}{}\right) \right)\right)
\nonumber \\
a\,b\;T^{\mu}_{||{\mu} \;8}
&=&
0\,=
3 a{} \left(b^{(0,1)}{} \left( \varphi ^{(0,1)}{}^2 \right)-b^{(1,0)}{} \varphi ^{(0,1)}{} \varphi ^{(1,0)}{} \right)
\nonumber \\
&-&
b{} \left( -3 a^{(0,1)}{} \varphi ^{(0,1)}{}^2+3 a^{(1,0)}{} \varphi ^{(1,0)}{} \varphi ^{(0,1)}{}
\right.
\nonumber \\
&-&
\left.
a{} \left( \varphi ^{(0,1)}{} \left(\, \varphi ^{(0,2)}{}-\varphi ^{(2,0)}{}\right) \right)\right)
\nonumber \\
.
\end{eqnarray}

Let
$
L_1 = \frac{\mathbf{a}^{(1,0)}(x^4,x^8)}{\mathbf{a}(x^4,x^8)}
$,
$
L_3 = \frac{\mathbf{a}^{(0,1)}(x^4,x^8)}{\mathbf{a}(x^4,x^8)}
$,
$
L_2 = \frac{\mathbf{b}^{(1,0)}(x^4,x^8)}{\mathbf{b}(x^4,x^8)}
$
and
$
L_4 = \frac{\mathbf{b}^{(0,1)}(x^4,x^8)}{\mathbf{b}(x^4,x^8)}
$.
The Euler-Lagrange equation for the inflaton field yields
\begin{eqnarray}
\label{ddphi}
\varphi ^{(2,0)}(x^4,x^8)&-&\varphi ^{(0,2)}(x^4,x^8)
+ 3 \varphi ^{(1,0)}(x^4,x^8)\left(L_1 + L_2 \right)
- 3 \varphi ^{(0,1)}(x^4,x^8)\left(L_3 + L_4 \right)
\nonumber \\
&=& 0
,
\end{eqnarray}
or, equivalently,
\begin{eqnarray}
\label{ddphia}
\varphi ^{(2,0)}(x^4,x^8)&+&
3 \varphi ^{(1,0)}(x^4,x^8)\left(L_1 + L_2 \right)
+ {\mu}^2 \, \varphi(x^4,x^8)
\nonumber \\
&=&
\varphi ^{(0,2)}(x^4,x^8)
+ 3 \varphi ^{(0,1)}(x^4,x^8)\left(L_3 + L_4 \right)
+ {\mu}^2 \, \varphi(x^4,x^8),
\end{eqnarray}
where ${\mu}^2$ is arbitrary.

Lastly, the field equations demand that the constraint equation
\begin{eqnarray}
\label{firstintegral}
L_1^2+3 L_1 L_2+L_2^2 \, &-& \, \frac{4}{3} \pi  \mathbb{G} \, \varphi ^{(1,0)}(x^4,x^8)^2
\nonumber \\
-
\left[
L_3^2+3 L_3 L_4+L_4^2 \,\right.&-&\left. \, \frac{4}{3} \pi  \mathbb{G} \, \varphi ^{(0,1)}(x^4,x^8)^2
\right]
\;=\; \frac{2 }{9} \Lambda
,
\end{eqnarray}
be satisfied.

To solve the field equations we use the fact that
$
\mathbb{P} = \ln{\left( a \,\times \,  b \right)}
$,
$
\mathbb{Q} = \ln{\left( \frac{b}{a} \right)}
$
and $\varphi$
satisfy, respectively, uncoupled, linear and linear field equations
of the form
\begin{eqnarray}
\label{wv}
&&f^{(0,2)}(  x^4  ,  x^8  ) - f ^{(2,0)}(  x^4  ,  x^8 )= \, S \,+\,
\nonumber \\
&\phantom{AWAW}&+\,3 \,\mathbb{P}^{(1,0)}(  x^4  ,  x^8  ) f ^{(1,0)}(  x^4  ,  x^8  )-3\, \mathbb{P}^{(0,1)}(  x^4  ,  x^8  ) f^{(0,1)}(  x^4  ,  x^8  )
,
\end{eqnarray}
where $f = {\mathbb{P}, \mathbb{Q} \textrm{  or  } \varphi}$ and $S = 0$ unless
$f = \mathbb{P}$ in which case  $S = -\frac{2}{3}  \Lambda $.
General solutions to these equations are substituted into
the constraint Eq[\ref{firstintegral}]
and the remaining field equations, which are then solved;
this procedure yields the solutions in the classes $E_0$ and $E_1$ that are described below.

\section{Class $E_0$ solution: Exact temporal exponential inflation/deflation}\label{sec:C0}

Using the technique described above we find that this model admits the following solutions,
which are exponential in $x^4$, periodic in  $x^8$ and
that are parameterized by $\Lambda$:
The scale factors are
\begin{eqnarray}
\label{yxx1}
a
&=&
{a(x^{4},x^{8})}
=
a_0
e^{\pm\,\frac{1}{3 }\sqrt{\frac{\Lambda } {2}} \, x^4 } \sqrt[12]{\sin ^2 \left(\, x^8 \,\sqrt{2\,\Lambda } \,\right)}
\nonumber \\
b
&=&
{b(x^{4},x^{8})}
=
b_0
e^{\mp\,\frac{1}{3 }\sqrt{\frac{\Lambda } {2}} \, x^4 } \sqrt[12]{\sin ^2 \left(\, x^8 \,\sqrt{2\,\Lambda } \,\right)}
,
\end{eqnarray}
where $a_0 $ and $b_0 $ are constants.
The relative expansion rate and the inflaton field are given by
\begin{eqnarray}
\label{yyxx1}
\ell_0
&=&
\mp\, \frac{1}{3 }\sqrt{\frac{\Lambda } {2}}
\nonumber \\
\varphi
&=&
\pm
\frac{1}{2} \sqrt{\frac{5}{6}} \ln \left[\tan ^2\left(\frac{1}{2} \sqrt{2 \Lambda }\, x^8 \,\right)\right]
.
\end{eqnarray}

For this case the scale factors $(a, b)$ have a spatial period equal to $\frac{1}{2}\; {C_8}$,
while the three functions  $(a, b, \varphi)$ possess a common spatial periodicity in the  $x^8$ coordinate
whose nonzero minimum value is
equal to the $x^8$-dimension spatial period
\begin{equation}
C_{8} = 2\, \pi \,R_8 =  \pi \sqrt{\frac{2}{ \Lambda }} =\,  \frac{\pi}{3 |\ell_0|},
\end{equation}

\begin{equation}\Lambda = \Lambda_0 =   \frac{1}{2}  \frac{1}{{R_{8}}^2}\end{equation}
For this case $|\ell_0|$ coincides with the Hubble parameter $H$.

Since a universe with many macroscopic times is not observed,  one may identify the physical solution for inflation by  choosing the solution
with the $-$ sign in the equation for $b$.
This solution then predicts the exponential deflation, with respect to time, of the
scale factor $b$ associated with the three extra time dimensions.
This   coincides with the exponential inflation, with respect to time, of the
scale factor $a$ associated with the three observed spatial dimensions.

\section{Temperature history of inflation in the Class $E_0$ background}\label{sec:th}

The reduced volume element $d \Omega$ on the hypersurface $x^4 = x^4_0 = $ constant is
\begin{equation}\label{DV}
d \Omega =  d \Omega(x^{\alpha}) =   \left[ {\sqrt{\det{\left( \mathbf{g}  \right) } } }\right]_{x^4 = x^4_0} \; {d }^{7} \tau
\end{equation}
where
\begin{equation}
{d }^{7} \tau = |d x^1 \wedge d x^2 \wedge d x^3 \wedge d x^5 \wedge d x^6 \wedge d x^7 \wedge d x^8   |.
\end{equation}
For the class $E_0$ solution
\begin{equation}\label{DV0}
d \Omega =  {d }^{7} \tau\, \left| {\sin  \left(\, x^8 \,\sqrt{2\,\Lambda } \,\right)}\right|,
\end{equation}
which clearly does not inflate.
The fact that $d \Omega$ vanishes on a set of measure zero is discussed in the \emph{Conclusion}.

To compute the temperature history $T = T(x^4)$ during inflation
we assume that before standard model particles are created the
chemical potential of the universe is zero.
We also assume that the entropy density, energy density, and
pressure describe only $\Lambda$ and the inflaton $\varphi$ field,
and that their averages over $x^8$ are functions $s(T), \rho(T),$ and $p(T)$ of the temperature $T$ alone.

In analogy with the conventional definition of thermal equilibrium
[see, for example, reference \cite{weinberg2008cosmology}, p.150, Eq. (3.1.1)],
we  assume that the $x^8$-average entropy in a co-moving
volume element is fixed:
\begin{equation}\label{s0}
s(T)
\,
\left<
\frac{d \Omega}{{d}^{6} \tau}
\right>_{x^8}\,
=\,\int_{-\pi}^{\pi}\,s(T, x^8) \, \left| {\sin  \left(\, x^8 \,\sqrt{2\,\Lambda } \,\right)}\right| \;d x^8
\,=\, \textrm{ constant},
\end{equation}
where
$
{d }^{6}\tau = |d x^1 \wedge d x^2 \wedge d x^3 \wedge d x^5 \wedge d x^6 \wedge d x^7    |
$.
Differentiating Eq.[\ref{s0}] with respect to $x^4$ yields
$
\frac{d\phantom{x^4}}{d x^4}\,
\left[
s(T(x^4)) \frac{d \Omega}{{d}^{6} \tau}
\right]
$
=
$
\frac{d T(x^4)}{d x^4}\,s'(T)  \, \frac{d \Omega}{{d}^{6} \tau}\,+\,s(T)\,
\frac{d\phantom{x^4}}{d x^4}\,\left[
\frac{d \Omega}{{d}^{6} \tau}
\right]
$
=
$
\dot{T}(x^4)\,s'(T) \, \frac{d \Omega}{{d}^{6} \tau} \,+\,0\,=\,0
$.
Assuming that $s'(T)\,\neq\,0$ gives
\begin{equation}
\frac{d\phantom{x^4}}{d x^4}\,T(x^4) \,=\,0.
\end{equation}
For this case the universe does not cool during inflation.
Reheating \cite{PhysRevD.56.3258} is a non-issue.

\section{The problem of the stability of a massive complex scalar field propagating on the Class $E_0$ background}\label{sec:smsf}

If a massive complex scalar field $\psi$ satisfies a linear field equation,
as does the $\psi$ field under discussion, then a quantum fluctuation  $\delta \psi$ of $\psi$
satisfies a linear field equation that is of the same form.
Therefore the question of the possible existence of instabilities of the massive complex scalar $\psi$ field that
are sourced by the momenta $\vec{w}$ associated to the extra time dimensions,
Issue [\ref{sss2}] raised in Section[\ref{sec:introduction}], may be investigated
by studying the propagation of
the  complex scalar  field $\psi$ on the background gravitational field,
but which is not coupled to the inflaton field $\varphi$.
In this paper the field equation for this scalar field
is given in Eq.[\ref{escact}].
Substituting the background gravitational field of Eq.[\ref{yxx1}]  into Eq.[\ref{escact}]
and using Eq.[\ref{spsi}] yields
\begin{eqnarray}
\label{escact00}
0
&=&
-\Psi ^{(2,0)}(\,x^4\,,\,x^8\,)
+\Psi ^{(0,2)}(\,x^4\,,\,x^8\,)
+
\sqrt{2\,\Lambda } \cot \left( \sqrt{2\,\Lambda } \,x^8\,\right) \Psi ^{(0,1)}(\,x^4\,,\,x^8\,)
\nonumber \\
&+&
\Lambda
\left(w^2 e^{\frac{1}{3}  \sqrt{2\,\Lambda } \,x^4\,}-k^2\,e^{-\frac{1}{3}  \sqrt{2\,\Lambda } \,x^4\,} \right)
\sqrt[6]{\csc^2 \left( \sqrt{2\,\Lambda } \,x^8\,\right)}
\Psi (\,x^4\,,\,x^8\,)
\nonumber \\
&-&
\Lambda
\left[  m^2\,+\,
\frac{1}{3}
\zeta
\left(
8+5 \csc ^2\left(\sqrt{2\,\Lambda } \,x^8\,\right)
\right)
\right] \Psi (\,x^4\,,\,x^8\,),
\end{eqnarray}
which is a non-separable linear wave equation.
Here
$\vec{k} = \left(k_1, k_2, k_3\right)^{T}\;$,
$\vec{w} = \left(k_5, k_6, k_7\right)^{T}$,
$
\Lambda\, k^2 \, =\vec{k} \cdot  \vec{k} = \,{k_1}^{2}+{k_2}^{2}+{k_3}^{2}\;
$
and
$
\Lambda\, w^2 \,=\vec{w} \cdot  \vec{w} = \, {k_5}^{2}+{k_6}^{2}+{k_7}^{2}
$;
we have also used the fact that the Ricci scalar is
$
\frac{1}{3} \Lambda  \left(5 \csc ^2\left(\sqrt{2\,\Lambda } \,x^8\,\right)+8\right)
$
for the class $E_0$  background gravitational field.
In Eq.[\ref{escact00}] we put
$
\Psi(x^4, x^8) =\psi\left(\sqrt{2\,\Lambda } \,x^4\,,\sqrt{2\,\Lambda } \,x^8\,\right)
$,
and then for notational simplicity  set
$
x = \sqrt{2\,\Lambda } \,x^8
$
(where $-\pi \leq x \leq \pi$),
and
$
t = \sqrt{2\,\Lambda } \,x^4\,
$.
Since $\Lambda = \frac{1}{2}$ for this case, $x = x^8$.
This yields a wave equation for the massive complex scalar $\psi (\,t\,,\,x\,)$ field that is given by
\begin{eqnarray}
\label{escact01}
0
&=&
-\psi ^{(2,0)}(\,t\,,\,x\,)
+\psi ^{(0,2)}(\,t\,,\,x\,)
+
\cot \left( x \right) \psi ^{(0,1)}(\,t\,,\,x\,)
\nonumber \\
&+&
\frac{1}{2}
\left(w^2 e^{\frac{1}{3} t\,}-k^2\,e^{-\frac{1}{3}t\,} \right)
\sqrt[6]{\csc^2 \left( x \right)}\;
\psi (\,t\,,\,x\,)
\nonumber \\
&-&
\frac{1}{2}
\left[  m^2\,+\,
\frac{1}{3}
\zeta
\left(
8+5 \csc ^2\left(x\right)
\right)
\right]
\psi (\,t\,,\,x\,).
\end{eqnarray}
Note that
\begin{equation}
\label{sh}
\frac{1}{2}\left( \, w^2\,e^{t/3} \,- \,k^2\, e^{-t/3}\right)
\,=\,
k \, w \,
\frac{1}{2}\left( \, \frac{w}{k}\,e^{t/3} \,- \,\frac{k}{w}\, e^{-t/3}\right)
\,=\,
k \, w \, \sinh \left[\frac{t}{3} + \ln \left(\frac{w}{k}\right)\right].
\end{equation}

Eq.[\ref{escact01}] is non-separable. Obviously one cannot write $\psi (\,t\,,\,x\,) = T (\,t\,)\,X (\,x\,)$ then
separate functions of $x$ from functions of $t$.
On the other hand, for a separable linear partial differential equation one
may separate variables, then introduce a separation constant
and in principle solve the ODE for eigenmodes, subject to given boundary conditions, and deduce a
dispersion relation that directly
relates a physical frequency $\omega $ to, say, the momentum wave vectors $(\vec{k}, \vec{w})$.
Eq.[\ref{escact01}] does not admit a simple dispersion relation that
relates a physical-frequency $\omega $ to the momentum wave vectors $(\vec{k}, \vec{w})$.

We study two special cases of Eq.[\ref{escact01}], the first with
$\zeta = 3/10$
and the second with
$\zeta = 0$.
The solution for $\zeta = 3/10$ is expanded into a series of Fourier modes
$\Sigma_{n_{8}=-\infty}^{\infty} f_{n_{8}}(t) \; \exp (i \,n_{8}\, x)$.
The second solution with
$\zeta = 0$ is expanded into  a series of (two) Legendre polynomial modes, each of the form
$
\sum_{{n_8}=0}^{\infty}\,h_{n_8}(t)\,\left[\sqrt{\left({n_8}+\frac{1}{2}\right)}P_{n_8}(\cos{x})\right]
$.
There is an expansion for each of the intervals
$-\pi \leq x \leq 0$ and  $0 \leq x \leq \pi$.
The two Legendre expansions must be matched at $x = 0$ and $x = \pm \pi$.

\subsubsection{Wave equation  with  $\zeta = 3/10$ }\label{sec:sss1}
We take advantage of the
coupling of the massive scalar field
to the Ricci scalar to simplify the wave equation.
In Eq.[\ref{escact01}] we first put
$
\psi(t, x) =\sin ^{-\frac{1}{2}}\left(x\right) F\left(t,x\right)
$.
This yields  \begin{eqnarray}
\label{sc00x}
0
&=&
F^{(2,0)}(t,x)-F^{(0,2)}(t,x)
\nonumber \\
&+&
\frac{1}{2}\left( \,k^2\, e^{-t/3}\,- \,w^2\, e^{t/3} \right) \,\sqrt[6]{\csc^2 (x)} \,F(t,x)
\nonumber \\
&+&
\frac{1}{12} \left(16 \zeta +6 m^2-3\right) F(t,x)+\frac{1}{12} (10 \zeta -3) \csc ^2(x) F(t,x).
\end{eqnarray}
In order to eliminate the
$\frac{1}{12} (10 \zeta -3)\csc ^2(x) F(t,x)$
term from this equation
we henceforth assume in this section that
$\zeta = 3/10$.

Substituting a Fourier decomposition\\
$F(t, x) = \Sigma_{n_{8}=-\infty}^{\infty} f_{n_{8}}(t) \; \exp (i \,n_{8}\, x)$
into Eq.[\ref{sc00x}],
with $\zeta = 3/10$,
and then multiplying by
$\frac{1}{2 \pi}\,\exp (-i \,m_{8}\, x)\;dx$
and integrating from  $(-\pi, \pi)$   yields
\begin{eqnarray}
\label{sc00Fx}
0&=&
\frac{d^2\phantom{t}}{d t^2}\,f_{m_{8}}(t)+
{m_{\textrm{eff}}}^2 \; f_{m_{8}}(t)
-
\frac{1}{2}\left( \, w^2\,e^{t/3} \,- \,k^2\, e^{-t/3}\right)
\sum _{n_{8}\,=\,-\infty }^{\infty }
\left\{c_{m_{8}\,n_{8}}
\,
f_{n_{8}}(t)
\,\right\}
\nonumber \\
&=&
\frac{d^2\phantom{t}}{d t^2}\,f_{m_{8}}(t)+
{m_{\textrm{eff}}}^2 \; f_{m_{8}}(t)
-\,
k \, w \, \sinh \left[\frac{t}{3} + \ln \left(\frac{w}{k}\right)\right]
\sum _{n_{8}\,=\,-\infty }^{\infty }
\left\{c_{m_{8}\,n_{8}}
\,
f_{n_{8}}(t)
\,\right\}
,
\end{eqnarray}
where
the  $c_{m_{8},n_{8}}$  are defined in Eq.[\ref{cnm}]
and
\begin{equation}
\label{effmassF}
{m_{\textrm{eff}}}^2
\,=\,
\frac{m^2}{2}+\frac{3}{20}+\,m_{8}\,^2
\end{equation}

As expected, the square of the effective mass ${m_{\textrm{eff}}}$
of the scalar field receives contributions from the spatial $x^8$ Fourier modes.
The square of the effective mass ${m_{\textrm{eff}}}$ also receives a contribution  $\zeta = 3/10$ from the
coupling with the Ricci scalar.

The matrix elements  $c_{m_{8},n_{8}}$  of $\sqrt[6]{\csc^2 (x)}$ in the Fourier basis are
\begin{eqnarray}\label{cnm}
c_{m_{8},n_{8}} &=&
\frac{1}{2 \pi }
\int_{-\pi}^{\pi}\; \sqrt[6]{\csc^2 (x)} \; \exp [i \left( n_{8} - m_{8}   \right) x] \; d x
\nonumber \\
&=&
\sqrt[3]{2}\;
\frac{ \Gamma \! \left(\frac{2}{3}\right)}{\Gamma\!  \left(\frac{5}{6}\right)^2}
\left\{ \begin{array}{ll}
(-1)^{N}\frac{{\Gamma\!  \left(\frac{5}{6}\right)^2}}{\Gamma \left(\frac{5}{6}-N\right) \Gamma \left(\frac{5}{6}+N\right)} & \textrm{ if }\, n_8\,-\,m_8 \,=\, 2\,N\,\textrm{ is even }\\
0 & \textrm{ if }\, n_8\,-\,m_8 \,\textrm{ is odd }
\end{array}
\right. .
\end{eqnarray}

We note that the $(-1)^{N}\frac{{\Gamma\!  \left(\frac{5}{6}\right)^2}}{\Gamma \left(\frac{5}{6}-N\right) \Gamma \left(\frac{5}{6}+N\right)} $ are rational numbers,
\begin{eqnarray}\label{gam}
\gamma_N&=&(-1)^{N}\frac{{\Gamma\!  \left(\frac{5}{6}\right)^2}}{\Gamma \left(\frac{5}{6}-N\right) \Gamma \left(\frac{5}{6}+N\right)}
\nonumber \\
&=&
\left\{1,\frac{1}{5},\frac{7}{55},\frac{91}{935},
\frac{1729}{21505},
\frac{8645}{124729},
\frac{7657}{124729},
\frac{283309}{5113889},
\frac{12182287}{240352783},
\right.
\nonumber \\
&&
\left.
\frac{596932063}{12738697499},
\frac{2984660315}{68325741131},
\frac{2800988911}{68325741131},
\frac{187666257037}{4851127620301},
\right.
\nonumber \\
&&
\left.
\frac{1957090966243}{53362403823311},
\frac{154610186333197}{4429079517334813},
\frac{773050931665985}{23187533943694021}
\ldots \right\}
\end{eqnarray}
for $N = 0, 1, 2,\ldots, 15, \ldots$.
The $c_{m_{8},n_{8}}$ are the matrix elements of a Toeplitz matrix
and verify
$c_{m_{8},n_{8}}\,=\,c_{m_{8}-n_{8},0}\,=\,c_{0, n_{8}-m_{8}}\;$.
Accordingly, Eq.[\ref{sc00Fx}] may be written as

\begin{eqnarray}
\label{sc00Ftjr}
0&=&
\frac{d^2}{d\,t^2}f_{m_{8}}(t)
+
\left\{
{m_{\textrm{eff}}}^2
-
\,\,
k \, w \, \sinh \left[\frac{t}{3} + \ln \left(\frac{w}{k}\right)\right]
c_{0,\,0}
\right\}
f_{m_{8}}(t)
\nonumber \\
&-&\,
k \, w \, \sinh \left[\frac{t}{3} + \ln \left(\frac{w}{k}\right)\right]
\sum _{n_{8}\,=\,1 }^{\infty }
c_{2n_{8},\,0}
\left[
f_{2 n_{8}+m_{8}}(t)
\,+\,
f_{-2 n_{8}+m_{8}}(t)
\,\right]
\end{eqnarray}

or
\newpage
\begin{mdframed}
\begin{eqnarray}
\label{mainFnEQ}
0&=&\,
\frac{d^2}{d\,t^2}f_{m_{8}}(t)
+
\left\{
{m_{\textrm{eff}}}^2
-
\,\,
\sqrt{Y}\;
2\,\sinh\left(\tau \right)
\right\}
f_{m_{8}}(t)
\nonumber \\
&-&\,
\sqrt{Y}\,
2\,\sinh\left(\tau\right)
\sum _{n_{8}\,=\,1 }^{\infty }
(-1)^{n_{8}}\frac{{\Gamma\!  \left(\frac{5}{6}\right)^2}}{\Gamma \left(\frac{5}{6}-n_{8}\right) \Gamma \left(\frac{5}{6}+n_{8}\right)}
\left[
f_{2 n_{8}+m_{8}}(t)
\,+\,
f_{-2 n_{8}+m_{8}}(t)
\,\right],
\nonumber \\
\end{eqnarray}
\end{mdframed}

where

\begin{equation}
\label{sTa}
\tau = \tau\left(t\right)=\frac{1}{3} t +  \log \left(\frac{w}{k}\right),
\end{equation}
\begin{equation}\label{Y}
\sqrt{Y} = \frac{1}{2}k \, w \,
\sqrt[3]{2}\;
\frac{ \Gamma \! \left(\frac{2}{3}\right)}{\Gamma\!  \left(\frac{5}{6}\right)^2}
\end{equation}
and
\begin{equation}\label{meff}
{m_{\textrm{eff}}}^2
\,=\,
\frac{m^2}{2}+\frac{3}{20}+\,m_{8}\,^2
.
\end{equation}

Because
$\frac{{\Gamma\!  \left(\frac{5}{6}\right)^2}}{\Gamma \left(\frac{5}{6}-N\right) \Gamma \left(\frac{5}{6}+N\right)} \propto
\left(\frac{1}{|N|}\right)^{2/3}$
as $N \rightarrow \pm\infty$,
the series expansions in  Eq.[\ref{sc00Fx}] and  Eq.[\ref{mainFnEQ}] exhibit a very long range coupling between modes.
In general the infinite sums in  Eq.[\ref{sc00Fx}]  and  Eq.[\ref{mainFnEQ}] will converge if $\left|f_{n_{8}}(t)\right| <
\left(\frac{1}{|n_{8}|}\right)^{1/3}$
as $|n_{8}| \rightarrow  \infty$.

In order to estimate the error of an approximate numerical solution to Eq.[\ref{mainFnEQ}] an  explicit approximate form for $\sqrt[6]{\csc^2 (x)} $
in terms of its Fourier components $c_{m_{8}\,n_{8}}$ is needed.
We find that
\begin{eqnarray}\label{csc3}
\sqrt[6]{\csc^2 (x)} &=&
\sum _{n_{8}\,=\,-\infty }^{\infty }
\left\{c_{m_{8}\,n_{8}}
\,
e^{ \dot{\imath} \,  n_{8} \, x}
\,\right\}
\nonumber \\
&=&
\frac{\sqrt[3]{2} \Gamma \left(\frac{2}{3}\right)}{\Gamma \left(\frac{5}{6}\right)^2}
{\sum _{{m_8}=-\infty}^{\infty} \,e^{2\,\dot{\imath} \, {m_8}\, x} \,
\left(
\frac{(-1)^{{m_8}}\, {\Gamma \left(\frac{5}{6}\right)}^2}{\Gamma \left(\frac{5}{6}-{m_8}\right) \Gamma \left(\frac{5}{6}+{m_8}\right)}
\right)
}
\nonumber \\
&=&
\frac{2 \sqrt{\frac{\pi }{3}}}{\Gamma \left(\frac{2}{3}\right) \Gamma \left(\frac{5}{6}\right)}
\left(
\, _2F_1\left(\frac{1}{6},1;\frac{5}{6};e^{2 i x}\right)+\, _2F_1\left(\frac{1}{6},1;\frac{5}{6};e^{-2 i x}\right)-1
\right)
\end{eqnarray}
almost everywhere, except $x = 0,\, \pm \,\pi$.
A plot of the difference\\
$
\sqrt[6]{\csc^2 (x)}
-
\frac{2 \sqrt{\frac{\pi }{3}}}{\Gamma \left(\frac{2}{3}\right) \Gamma \left(\frac{5}{6}\right)}
\left[
\, _2F_1\left(\frac{1}{6},1;\frac{5}{6};e^{2 i x}\right)+\, _2F_1\left(\frac{1}{6},1;\frac{5}{6};e^{-2 i x}\right)-1
\right]
$ \\
is given in Figure[1].
Let $n \in \mathbb{N}$, the natural numbers (excluding 0).
Given an approximate numerical solution that contains only a finite number of Fourier modes
$-n \leq n_8 \leq n$
one may make the substitution
based on Eq.[\ref{csc3}]
\begin{equation}\label{CSC3approx}
\sqrt[6]{\csc ^2(x)}
\rightarrow
\frac{\sqrt[3]{2} \Gamma \left(\frac{2}{3}\right)}{\Gamma \left(\frac{5}{6}\right)^2}
{\sum _{{m8}=-\left[\frac{n}{2}\right]}^{\left[\frac{n}{2}\right]} \frac{e^{i (2 {m8}) x} \left((-1)^{{m8}} \Gamma \left(\frac{5}{6}\right)^2\right)}{\Gamma \left(\frac{5}{6}-{m8}\right) \Gamma \left({m8}+\frac{5}{6}\right)}}
\end{equation}
when computing the numerical error.
Here, $\left[\frac{n}{2}\right]$ gives the greatest integer less than or equal to $\frac{n}{2}$.

We call the first line of Eq.[\ref{mainFnEQ}]
\begin{eqnarray}
\label{mainFnEQ1}
\mathcal{D}[f](\,k,\,w; \, m_8; \,t) &=&
\nonumber \\
\frac{d^2}{d\,t^2}f_{m_{8}}(t)
&+&
\left[
{m_{\textrm{eff}}}^2
-
\,
\frac{1}{2}\left( \, w^2\,e^{t/3} \,- \,k^2\, e^{-t/3}\right)\,
c_{0,\,0}
\right]
f_{m_{8}}(t)
\end{eqnarray}
the \emph{diagonal contribution} to Eq.[\ref{mainFnEQ}].
Here $c_{0,\,0} = \sqrt[3]{2} \; \frac{\Gamma \left(\frac{2}{3}\right)}{\Gamma \left(\frac{5}{6}\right)^2}$.

\subsubsection{Special solutions to  $\mathcal{D}[f](\,k,\,w; \, m_8; \,t) =0$ }\label{sec:dsss1}

$\mathcal{D}[f](\,k,\,w; \, m_8; \,t) =0$
possesses  simple mathematical solutions for the two special cases
$k = 0$ and $w = 0$,
which provide much  insight into this problem, but which are not physical
because the long range coupling with the other modes is completely neglected.
These solutions are discussed in the next section.

Let
$J_{n}\left(z\right)$ denote an ordinary Bessel function of the first type,
$Y_{n}\left(z\right)$ denote an ordinary Bessel function of the second type (Neumann function),
$I_{\lambda}\left(z\right)$ denote the modified Bessel function of the first type and order $\lambda$,
and
$K_{\lambda}\left(z\right)$ denote a modified Bessel function of the second type
(also known as the modified Bessel function of the third kind).

If $k=0$ then the general solution to $\mathcal{D}[f](\,k=0,\,w; \, m_8; \,t) =0$ is
\begin{equation}\label{scfk0}
f_{m_{8}}(t)=
c_1 \,K_{6\, \dot{\imath} \, {m_{\textrm{eff}}} }
\left( w\, e^{t/6}\, \alpha \,\right)
+
c_2 \,I_{6\, \dot{\imath} \, {m_{\textrm{eff}}} }
\left( w\, e^{t/6}\, \alpha \,\right),
\end{equation}
where
\begin{equation}\label{alpha}
\alpha\,=\,\sqrt{18\; c_{0,\,0}}
\,=\,3\,\times 2^{2/3}\,\frac{\sqrt{\Gamma \left(\frac{2}{3}\right)}}{\Gamma \left(\frac{5}{6}\right)}
\end{equation}
and
the $(c_1 , c_2)$ are arbitrary constants.
Asymptotically, for large argument $|z|$, the modified Bessel functions  behave as
$K_{\lambda}\left(z\right)  \sim e^{-z} \sqrt{\frac{\pi }{2 z}}+\cdots$
and
$I_{\lambda}\left(z\right)  \sim e^{z} \sqrt{\frac{ 1}{2 \pi z}}+\cdots$ \cite{MR0010746}.
Therefore $f_{m_{8}}(t)$ is a stable solution as $w \rightarrow\infty$ if  $c_2 = 0$.
Moreover,
$K_{\dot{\imath}\,\nu}\left(z\right)$
possesses a well known   integral representation
\begin{equation}
\label{Kirep}
K_{\dot{\imath}\,\nu}\left(z\right)
=
\int_{0}^{\infty}\cos{(\nu\, \xi)}\,{e}^{-z\, \cosh{(\xi)}}\,d\,\xi
,
\end{equation}
valid for $\arg{z} < \frac{\pi}{2}$,
which enables a simple demonstration  that $f_{m_{8}}(t)$ is stable as $w \rightarrow\infty$.

A stable solution to $\mathcal{D}[f](\,k=0,\,w; \, m_8; \,t) =0$ may be obtained
by choosing an initial condition for
$\frac{f_{m_{8}}'(t)}{f_{m_{8}}(t)}$
so that $c_2$ vanishes.
The required initial condition is
\begin{equation}\label{icf}
\frac{f_{m_{8}}'(0)}{f_{m_{8}}(0)}
=
\left\{\,\frac{\frac{\partial}{\partial\,t}K_{6\, \dot{\imath} \, {m_{\textrm{eff}}} }
\left( w\, e^{t/6}\, \alpha \,\right)}
{\,K_{6\, \dot{\imath} \, {m_{\textrm{eff}}} }
\left(  w\, e^{t/6}\, \alpha \,\right)}
\right\}_{t=0}\,
=
\left\{\frac{1}{6} \; W \; \frac{\partial \log (K_{6\, \dot{\imath} \, {m_{\textrm{eff}}} }(W))}{\partial W}\right\}_{\; W = \,w\, \alpha\;}
,
\end{equation}
where $\alpha$ is defined in Eq.[\ref{alpha}].
To make this solution also well-behaved as $w \rightarrow 0$ one can, for example,
put $c_1 = \frac{w}{1+w} C_1$.

If one  attempts to numerically integrate $\mathcal{D}[f](\,k=0,\,w; \, m_8; \,t) =0$
using a numerical quadrature algorithm then
at each time step $t_n$ of the process
one must maintain tight control of the error
\begin{equation}\label{error}
\left\{\frac{f_{m_{8}}'(t_n)}{f_{m_{8}}(t_n)}\right\}_{\textrm{numerical}}
-
\left\{\,\frac{\frac{\partial}{\partial\,t}K_{6\, \dot{\imath} \, {m_{\textrm{eff}}} }
\left( w\, e^{t/6}\, \alpha \,\right)}
{\,K_{6\, \dot{\imath} \, {m_{\textrm{eff}}} }
\left(  w\, e^{t/6}\, \alpha \,\right)}
\right\}_{t=t_n},
\end{equation}
because a non-zero value of this difference will source a
contribution from $\,I_{6\, \dot{\imath} \, {m_{\textrm{eff}}} }
\left( w\, e^{t/6}\, \alpha \,\right)$
in the numerical solution,
which may eventually exponentially overwhelm the true analytical solution.
This problem is stiff.

We note in passing that the behavior of the modified Bessel function of the second kind
with pure imaginary order, $K_{i \nu }(z)$, has been investigated
by Balogh \cite{Balogh:1965} and others.
Balogh has proved
that $K_{i \nu }(\frac{\nu}{p})$ is a positive monotone decreasing convex function of $p$
for $0 <p < 1$ and it  oscillates boundedly for $p > 1$, having a
countably infinite number of zeros.
Balogh gives
asymptotic expansions of the zeros of $K_{i \nu }(z)$ and  its derivative that are
uniform with respect to the enumeration of the zeros.

If $w=0$ then the stable solution to $\mathcal{D}[f](\,k,\,w=0; \, m_8; \,t) =0$ is
\begin{equation}\label{scfk1}
f_{m_{8}}(t)=
c_3 \,J_{6\, \dot{\imath} \, {m_{\textrm{eff}}} }
\left( k\, e^{-t/6}\, \alpha \,\right)
+
c_4 \,Y_{6\, \dot{\imath} \, {m_{\textrm{eff}}} }
\left( k\, e^{-t/6}\, \alpha \,\right),
\end{equation}
where the $(c_3 , c_4)$ are arbitrary constants.

\subsection{Wave equation  with  $\zeta = 0$}\label{sec:sss2}

We put $\zeta = 0$ in Eq.[\ref{escact01}] to obtain the wave equation for a minimally coupled
massive complex scalar field propagating through the background class $E_0$ gravitational field: \begin{eqnarray}
\label{escact02}
0
&=&
-\psi ^{(2,0)}(\,t\,,\,x\,)
+\psi ^{(0,2)}(\,t\,,\,x\,)
+
\cot \left( x \right) \psi ^{(0,1)}(\,t\,,\,x\,)
\nonumber \\
&-&
\frac{1}{2}m^2\,
\psi (\,t\,,\,x\,)
+
\frac{1}{2}
\left(w^2 e^{\frac{1}{3} t\,}-k^2\,e^{-\frac{1}{3}t\,} \right)
\sqrt[6]{\csc^2 \left( x \right)}\;
\psi (\,t\,,\,x\,)
\nonumber \\
&=&
-\psi ^{(2,0)}(\,t\,,\,x\,)
+
\frac{1}{\sin \left( x \right)}
\frac{\partial}{\partial\,x}
\left[
\sin \left( x \right) \frac{\partial}{\partial\,x}\psi(\,t\,,\,x\,)
\right]
\nonumber \\
&-&
\frac{1}{2}m^2\,
\psi (\,t\,,\,x\,)
+
\frac{1}{2}
\left(w^2 e^{\frac{1}{3} t\,}-k^2\,e^{-\frac{1}{3}t\,} \right)
\sqrt[6]{\csc^2 \left( x \right)}\;
\psi (\,t\,,\,x\,)
.
\end{eqnarray}

Recall that $-\pi \leq x \leq \pi$ and \emph{not}  $0 \leq x \leq \pi$.
In order to obtain a solution to Eq.[\ref{escact02}] we solve for  $\psi (\,t\,,\,x\,)$
on each of the two intervals
$\mathcal{I}_{\,-} = -\pi \leq x < 0$ and  $\mathcal{I}_{+} = 0 \leq x \leq \pi$
and then enforce appropriate continuity and boundary conditions:
We assume that $\psi (\,t\,,\,x\,)$
and
$\frac{\partial}{\partial\,x}\psi (\,t\,,\,x\,)$
are continuous,
and  that  $\psi (\,t\,,\,-\pi\,) = \psi (\,t\,,\,\pi\,)$,
since $x$ may be regarded as a canonical angular coordinate $\theta$ for the boundary of the unit circle.
Note that, with $X = \cos x= \cos \theta,\; x = \theta = \pm \arccos X$,  Eq.[\ref{escact02}] maps to
\begin{eqnarray}
&-&\psi ^{(2,0)}(t,X)+\left(1-X^2\right) \psi ^{(0,2)}(t,X)-2 X \psi ^{(0,1)}(t,X)-\frac{m^2}{2}\psi (t,X)
\nonumber \\
&+&
\frac{1}{2}\left( w^2 e^{t/3}-k^2 e^{-t/3}\right)\left( \sqrt[6]{\frac{1}{1-X^2}}\right) \psi (t,X)
= 0.
\end{eqnarray}
Let
$U(x)$ denote the unit step function
\begin{equation}
\label{ux}
U(x)
=
\left\{
\begin{array}{cc}
0 &\;\textrm{ for } x < 0 \\
\frac{1}{2} &\;\textrm{ for } x = 0 \\
1 &\;\textrm{ for } x > 0 \;;
\end{array}
\right.
\end{equation}
accordingly,
$\psi (\,t\,,\,x\,)$ may be expressed as
\begin{eqnarray}
\label{lp}
\psi (\,t\,,\,x\,)
&=&
\psi (\,t\,,\,x\,)\rfloor_{\mathcal{I}_{-}}
U(-x)
+
\psi (\,t\,,\,x\,)\rfloor_{\mathcal{I}_{+}}
U(x)
\nonumber \\
&=&
\psi_{-} (\,t\,,\,x\,)
U(-x)
+
\psi_{+} (\,t\,,\,x\,)
U(x)
,
\end{eqnarray}
and
$\frac{\partial}{\partial\,x}\psi (\,t\,,\,x\,)$ is given by
\begin{eqnarray}
\label{dlp}
\psi^{(0,1)}(\,t\,,\,x\,)
&\equiv&
\frac{\partial}{\partial\,x}
\psi (\,t\,,\,x\,)
\nonumber \\
&=&
U(-x)
\frac{\partial}{\partial\,x}\psi _{-}(\,t\,,\,x\,)
+
U(x)
\frac{\partial}{\partial\,x}\psi_{+} (\,t\,,\,x\,)
\nonumber \\
&\;&+\;
\delta(x)
\left[
\psi_{+} (\,t\,,\,0\,)
-
\psi_{-} (\,t\,,\,0\,)
\right]
.
\end{eqnarray}
As stated above, we assume that $\psi (\,t\,,\,x\,)$
and
$\frac{\partial}{\partial\,x}\psi (\,t\,,\,x\,)$
are continuous,
and  that  $\psi (\,t\,,\,-\pi\,) = \psi (\,t\,,\,\pi\,)$.
These conditions imply that
\begin{equation}
\label{plim}
\psi(\,t\,,\,0\,) \, = \,
\frac{1}{2}
\left[
\psi_{-} (\,t\,,\,0\,)+\psi_{+} (\,t\,,\,0\,)
\right]
=
\left\{
\begin{array}{c}
\lim_{x \rightarrow 0^+} \psi_{-} (\,t\,,\,x\,) \\
\lim_{x \rightarrow 0^-} \psi_{+} (\,t\,,\,x\,)\;
\end{array}
\right.,
\end{equation}
whence
\begin{equation}
\label{p0lim}
\psi_{-} (\,t\,,\,0\,)=\psi_{+} (\,t\,,\,0\,),
\end{equation}
and, using Eq.[\ref{p0lim}] in Eq.[\ref{dlp}],
\begin{equation}
\label{dplim}
\psi^{(0,1)}(\,t\,,\,0\,) \, = \,
\frac{1}{2}
\left[
\psi_{-}^{(0,1)} (\,t\,,\,0\,)+\psi_{+}^{(0,1)} (\,t\,,\,0\,)
\right]
=
\left\{
\begin{array}{c}
\lim_{x \rightarrow 0^+} \psi_{-}^{(0,1)} (\,t\,,\,x\,) \\
\lim_{x \rightarrow 0^-} \psi_{+}^{(0,1)} (\,t\,,\,x\,)\;.
\end{array}
\right.
\end{equation}
This implies that
\begin{equation}
\label{dp0lim}
\psi_{-}^{(0,1)} (\,t\,,\,0\,)=\psi_{+}^{(0,1)} (\,t\,,\,0\,)
\end{equation}

For simplicity, in order to satisfy the boundary and continuity conditions
we henceforth assume for the remainder of this paper that
\begin{equation}
\label{psim2}
\psi^{(0,1)}(\,t\,,\,0\,) \, = \,0\,=\,
\psi_{-}^{(0,1)} (\,t\,,\,0\,)=\psi_{+}^{(0,1)} (\,t\,,\,0\,)
\end{equation}
and that
when $x \in \mathcal{I}_{\,-}$,
$\;-\pi \leq x < 0$,
\begin{equation}
\label{psim}
\psi_{-} (\,t\,,\,x\,) \, = \,\psi_{+} (\,t\,,\,-x\,) \;\textrm{, }\;x \in \mathcal{I}_{\,-}
\;.
\end{equation}
Next, let us expand $\psi (\,t\,,\,x\,)$ on each of the two intervals
$\mathcal{I}_{\,-} = -\pi \leq x < 0$ and  $\mathcal{I}_{+} = 0 \leq x \leq \pi$
in terms of a series of Legendre polynomials of the form
\begin{equation}
\label{lpe}
\psi_{\pm} (\,t\,,\,x\,)
\equiv
\psi (\,t\,,\,x\,)\rfloor_{\mathcal{I}_{\pm}}
=
\sum_{{n_8}=0}^{\infty}\,h^{(\pm)}_{n_8}(t)\,\left[\sqrt{\left({n_8}+\frac{1}{2}\right)}P_{n_8}(\cos{x})\right]
.
\end{equation}
In virtue of  Eq.[\ref{psim}], $h^{(-)}_{n_8}(t) = h^{(+)}_{n_8}(t)$.
For brevity we put  $h^{(+)}_{n_8}(t) = h_{n_8}(t)$.
Before substituting Eq.[\ref{lpe}] into Eq.[\ref{escact02}] and simplifying we
record the following  definitions.

Let $\mathbb{N}$ denote  the non-negative integers ${0, 1, 2, 3, \ldots}\;$ .
Define, $ \forall \; m_8, n_8 \in \mathbb{N}$,
\begin{eqnarray}
\label{dmn}
{d}_{m_8\,n_8}&=&{d}_{m_8,\,n_8}
\nonumber \\
&=&
\sqrt{\left({m_8}+\frac{1}{2}\right)}
\sqrt{\left({n_8}+\frac{1}{2}\right)}
\int_{0}^{\pi}
P_{m_8}(\cos{x})
P_{n_8}(\cos{x})
\,\sin ^{\frac{2}{3}}(x)\,dx
\nonumber \\
&=&
\sqrt{\left({m_8}+\frac{1}{2}\right)}
\sqrt{\left({n_8}+\frac{1}{2}\right)}
\int_{-1}^{1}
P_{m_8}(X)
P_{n_8}(X)
\frac{1}{\sqrt[6]{1-X^2}}
\,dX
\nonumber \\
&=&
\pi  \Gamma \left(\frac{5}{6}\right)^2 \sqrt{\left( {{m_8}}+\frac{1}{2}\right) \left( {{n_8}}+\frac{1}{2}\right)}
\nonumber \\
&&\sum_{j=\left|  {{n_8}}- {{m_8}} \right|\textrm{, step 2} }^{ {m_8} +  {n_8}}
\left\{
\frac
{
(2 j+1) (-1)^{j+ {{m_8}}+ {{n_8}}}
}
{
\Gamma \left(\frac{1}{2}-\frac{j}{2}\right) \Gamma \left(\frac{5}{6}-\frac{j}{2}\right)
\Gamma \left(\frac{j}{2}+1\right) \Gamma \left(\frac{j}{2}+\frac{4}{3}\right)
}
\right.
\nonumber \\
&&
\left.
\frac
{
\Gamma (-j+ {{m_8}}+ {{n_8}}+1) \Gamma (j- {{m_8}}+ {{n_8}}+1)\Gamma (j+ {{m_8}}- {{n_8}}+1)
}
{
\Gamma (j+ {{m_8}}+ {{n_8}}+2)
}
\right.
\nonumber \\
&&
\left.
\left[
\frac
{
\Gamma \left(\frac{1}{2} (j+ {{m_8}}+ {{n_8}}+2)\right)
}
{
\Gamma \left(\frac{1}{2} (-j+ {{m_8}}+ {{n_8}}+2)\right) \Gamma \left(\frac{1}{2} (j- {{m_8}}+ {{n_8}}+2)\right) \Gamma \left(\frac{1}{2} (j+ {{m_8}}- {{n_8}}+2)\right)
}
\right]^2
\right\}
\nonumber \\
\end{eqnarray}
(of course, $\sin ^{\frac{2}{3}}(x)=\sqrt[6]{\csc^2 \left( x \right)}\;\sin{x}$ on $0\leq x\leq\pi$).
One observes that
\begin{equation}
{d}_{2 m_8+1,\, 2 n_8}=0={d}_{2 m_8,\, 2 n_8+1}\; \forall \; m_8, n_8 \in \mathbb{N}.
\end{equation}
Substituting Eq.[\ref{lpe}] into Eq.[\ref{escact02}], multiplying by
$\left[\sqrt{\left({m_8}+\frac{1}{2}\right)}P_{m_8}(\cos{x})\right]\,\sin{x}\,dx$,
$m_8 \in \mathbb{N}$,
and integrating over the
interval $0 \leq x \leq \pi$ yields
\begin{equation}
\label{escact02yn}
\ddot{h}_{m_8}(t)+
\left[
\frac{1}{2}m^2\,+{m_8}({m_8}+1)
\right]
{h}_{m_8}(t)
+
\frac{1}{2}
\left(k^2\,e^{-\frac{1}{3}t}- w^2 e^{\frac{1}{3} t\,}\right)
\sum_{{n_8}=0}^{\infty}\,
{d}_{m_8\,n_8}{h}_{n_8}(t)=0.
\end{equation}

For this case the effective mass ${m_{\textrm{eff}}}$ is given by
\begin{equation}
\label{effmassL}
{m_{\textrm{eff}}}^2
\,=\,\frac{1}{2}m^2\,+\,{m_8}({m_8}+1)
\end{equation}

We find that
${d}_{m_8\,n_8} \propto
\left(\frac{1}{n_{8}}\right)^{2/3}$
for
$n_{8} >> m_{8}$
as $n_{8} \rightarrow \pm\infty$.
the series expansion in  Eq.[\ref{escact02yn}] also exhibits a very long range coupling between modes.
The infinite sum in  Eq.[\ref{escact02yn}]  will converge if $\left|{h}_{n_8}(t)\right| <
\left(\frac{1}{n_{8}}\right)^{1/3}$
as $n_{8} \rightarrow \pm\infty$.

The diagonal contribution to Eq.[\ref{escact02yn}] is
\begin{eqnarray}
\label{escact02yn1}
\mathcal{D}[h](\,k,\,w; \, m_8; \,t) &=&
\nonumber \\
\ddot{h}_{m_8}(t)+
\left[
{m_{\textrm{eff}}}^2
\right.
&-&
\left.
\,
\frac{1}{2}\left( \, w^2\,e^{t/3} \,- \,k^2\, e^{-t/3}\right)\,
{d}_{m_8\,m_8}
\right]
{h}_{m_8}(t)
,
\end{eqnarray}
which has the same form as
the diagonal contribution to Eq.[\ref{mainFnEQ}].

Setting to zero the diagonal contribution to either Eq.[\ref{mainFnEQ}] or  Eq.[\ref{escact02yn}] yields an equation
of the form

\begin{eqnarray}
\label{modeD1}
-
\ddot{Y}_{m_8}(t)
\,&+&\,
B\,\frac{1}{2}
\left( \, w^2\,e^{t/3} \,- \,k^2\, e^{-t/3}\right)\,
{Y}_{m_8}(t)
\,=\,
{m_{\textrm{eff}}}^{2}\,
{Y}_{m_8}(t)
,\\
\textrm{ where }
&&
\nonumber \\
&&
\left\{
\begin{array}{lll}
{m_{\textrm{eff}}}^2=\,\frac{3}{20}+\frac{1}{2}m^2\,+{m_8}^2,  &
B=\, c_{0,0},  &\textrm{ for }
\zeta = \frac{3}{10}, \\
{m_{\textrm{eff}}}^2=\frac{1}{2}m^2\,+{m_8}({m_8}+1), &
B=B(m_8)=\, {d}_{m_8\,m_8}, &\textrm{ for }
\zeta = 0
\end{array}
\right.
\label{modeD2}
\nonumber \\
\end{eqnarray}

\section{Existence of stable solutions }\label{sec:sswhy}
We recall the content of Issue [\ref{sss2}],
\emph{Momenta corresponding to the extra time dimensions induce exponentially rapid growth of quantum fluctuations of the field; the universe is unstable.
This instability is associated with the very largest momenta (shortest wavelengths)}.

In this paper we are concerned with
exponentially rapid growth of \emph{quantum fluctuations of the field}. This causes perturbation theory to break down since
$| \delta \psi |$ becomes of order one faster than some positive power of $t$.  Exponentially rapid growth of the field $\psi$ itself is not an instability. Instability of the field $\psi$
means  the norm $| \psi |$ of the field goes to infinity in finite time. This is not the issue under discussion.
``Stable" means that
solutions $f_{m_8} = \delta \psi_{m_8}$ of Eq.[\ref{mainFnEQ}]  and   $h_{m_8} = \delta \psi_{m_8}$ of Eq.[\ref{escact02yn}] exist
that possess a norm that is non-exponentially increasing as $w$ increases.  Then if $| \delta \psi |$ is initially small then
$| \delta \psi |$ does not become of order one exponentially with time.

A formal argument can be made  that demonstrates that
\emph{stable solutions}  $f_{m_8} = \delta \psi_{m_8}$ to Eq.[\ref{mainFnEQ}]  and   $h_{m_8} = \delta \psi_{m_8}$ to Eq.[\ref{escact02yn}] always exist.
Since both  Eq.[\ref{mainFnEQ}]  and  Eq.[\ref{escact02yn}] have similar forms,
our discussion will focus on   Eq.[\ref{mainFnEQ}];
our conclusions will be valid for both equations.

The phrase \emph{stable solution} means, most importantly, that the norm of the solution
is finite, for fixed time $t$ and $k$,
as $w \rightarrow \infty $
[recall that
$\vec{k} = \left(k_1, k_2, k_3\right)^{T}\;$,
$\vec{w} = \left(k_5, k_6, k_7\right)^{T}$,
$
\Lambda\, k^2 \, =\vec{k} \cdot  \vec{k} = \,{k_1}^{2}+{k_2}^{2}+{k_3}^{2}\;
$
and
$
\Lambda\, w^2 \,=\vec{w} \cdot  \vec{w} = \, {k_5}^{2}+{k_6}^{2}+{k_7}^{2}
$].

The argument begins as follows. We affinely transform the physical time $t$ to
a new time coordinate $\tau$ that is defined by setting
\begin{equation}
\label{st}
t=3 \tau + 3 \log \left(\frac{k}{w}\right)
\end{equation}
As we have seen above, under this transformation
\begin{eqnarray}
\label{shT}
&&\frac{1}{2}\left( \, w^2\,e^{t/3} \,- \,k^2\, e^{-t/3}\right)
\,=\,
k \, w \,
\frac{1}{2}\left( \, \frac{w}{k}\,e^{t/3} \,- \,\frac{k}{w}\, e^{-t/3}\right)
\nonumber \\
\,&=&\,
k \, w \, \sinh \left[\frac{t}{3} + \ln \left(\frac{w}{k}\right)\right]
\,=\, k\,w\,\sinh{\left(\tau\right)}.
\end{eqnarray}

In terms of the new time coordinate $\tau$,  $\psi = \psi(\tau,x)$ satisfies the wave equation Eq.[\ref{sc00xeq}], below.
Since
the quantum fluctuations of $\psi$ satisfy a field equation similar to Eq.[\ref{sc00xeq}],
we limit our discussion of stability to solutions of Eq.[\ref{sc00xeq}]
and its relatives.

The inverse transformation to the map of the time coordinate $t$ defined in Eq.[\ref{st}] is
\begin{equation}
\label{sT}
\tau=\frac{1}{3} t +  \log \left(\frac{w}{k}\right).
\end{equation}
For fixed $t$ and  $k$, as $w \rightarrow \infty$ then $\tau \rightarrow \infty$.
For fixed $t$ and  $k$ a solution $ \delta \psi(\tau,x)$ is called \emph{stable} if the norm $\| \delta \psi\|$ of the solution
to the wave equation Eq.[\ref{sc00xeq}], below,
satisfies $0 \leq \| \delta \psi\| < \infty$ as $\tau \rightarrow \infty$.
Note that an effective frequency-like parameter $\varpi$, with
${{{\varpi}^{2}}} \,=\,2\, \sqrt{Y}\,
\propto \, k\, w$
appears in  Eq.[\ref{mainHnEQ}] and  Eq.[\ref{sc00xeq}], below.
For  fixed $w$, as $k \rightarrow \infty$ then $\varpi \rightarrow \infty$
and
for  fixed $k$, as $w \rightarrow \infty$ then $\varpi \rightarrow \infty$.
The dependence of  $ \delta \psi(\tau,x)$  on  $\varpi$
as $\varpi \rightarrow \infty$ does not enter into our definition of stability.
$ \delta \psi(\tau,x)$ oscillates at higher and higher frequencies as  $\varpi \rightarrow \infty$,
but does not become unstable.
Both non-resonant and resonant oscillations do not drive
$| \delta \psi |$ to order one faster than some positive power $t^{\gamma}$ of $t$.

Under the change of time coordinate defined  in Eq.[\ref{st}], Eq.[\ref{mainFnEQ}] maps to
\\
\begin{mdframed}
\begin{eqnarray}
\label{mainHnEQ}
0&=&
\frac{1}{9} \;
\frac{d^2\phantom{\tau}}{d \tau^2}\,
h_{n_8}(\tau)+
\left\{
{m_{\textrm{eff}}}^2\,
-
\,
\sqrt{Y}\;
2\,\sinh\left(\tau \right)
\right\}
h_{n_8}(\tau)
\nonumber \\
&-&\,
\sqrt{Y}\,
2\,\sinh\left(\tau\right)
\sum _{m_{8}\,=\,1 }^{\infty }
(-1)^{m_{8}}\frac{{\Gamma\!  \left(\frac{5}{6}\right)^2}}{\Gamma \left(\frac{5}{6}-m_{8}\right) \Gamma \left(\frac{5}{6}+m_{8}\right)}
\left[
h_{2 m_{8}+n_{8}}(\tau)
\,+\,
h_{-2 m_{8}+n_{8}}(\tau)
\,\right],
\nonumber \\
\end{eqnarray}
\end{mdframed}
where
$h_{n_8}(\tau) = f_{n_8}(t) = \delta \psi_{n_8}(t)$,
${m_{\textrm{eff}}}^2
\,=\,
\frac{m^2}{2}+\frac{3}{20}+\,n_{8}\,^2$
(see Eq.[\ref{meff}])
and
$\sqrt{Y} = \frac{1}{2}k \, w \,
\sqrt[3]{2}\;
\frac{ \Gamma \! \left(\frac{2}{3}\right)}{\Gamma\!  \left(\frac{5}{6}\right)^2}$
(see Eq.[\ref{Y}]);
the independent `modes' are labelled by $n_8$.
From the form of this equation one sees that only the product
$k\, w$, and not $k$ and $w$ independently, govern the behavior of the solutions.
Note that Eq.[\ref{mainHnEQ}] has the property that
the even modes and odd modes evolve independently of each other
(because $ \,\sqrt[6]{\csc^2 (x)} \,$ is an even function of $x$).

According to the theory of systems of linear differential equations
a solution associated to Eq.[\ref{mainHnEQ}] exists that is expressible in the form
\begin{eqnarray}
\label{sc00FThn8}
h_{n_8}(\tau)&=&
\left\{
\begin{array}{cc}
\sum_{j=-\infty}^{\infty}
h_{n_{8}\,; j} \; \exp{ \left(\left(-\sqrt{a_2}\,+\, j\right)\, \tau \;\right)}&\textrm{for  } n_8 \textrm{  even}\\
\sum_{j=-\infty}^{\infty}
h_{n_{8}\,; j} \; \exp \left(\left(-\sqrt{a_1}\,+\, j\right)\, \tau \;\right)&\textrm{for  } n_8 \textrm{  odd}
\end{array}
\right.
,
\end{eqnarray}
where the $h_{n_{8}\,; j}, a_1, a_2$ are constants to be determined (see, for example, Ref.\cite{dougall1915solution}
Equation[7], page 177).
We substitute
Eq.[\ref{sc00FThn8}] into  Eq.[\ref{mainHnEQ}] and then
demand that the coefficient of each
$ \; e^{j \, \tau },\;\;  j = -\infty \ldots\, \infty$, vanish,
where $a = a_1$ if $n_8$ is odd and  $a = a_2$ if $n_8$ is even.
This yields the  system of linear equations  for
the   $h_{n_{8}\,; j} \;$ given by
\begin{eqnarray}
\label{sc00heq}
&0&=
-h_{n_{8}\,; j} \; \left[\frac{3}{20}+\frac{m^2}{2}+ {n_8}^2+\frac{1}{9} \left(j-\sqrt{ {a}}\right)^2\right]
+\sqrt{Y} \left(h_{n_{8}\,; j-1} \; - \, h_{n_{8}\,; j+1} \;\right)
\nonumber \\
&+&
\sqrt{Y} \sum _{ {m_8}=1}^{\infty } \frac{(-1)^{ {m_8}} \Gamma \left(\frac{5}{6}\right)^2 }{\Gamma \left(\frac{5}{6}- {m_8}\right) \Gamma \left( {\frac{5}{6}+m_8}\right)}
\left(h_{n_{8}-2 {m_8}\,  ; j-1}\,-h_{n_{8}-2 {m_8}\,\,; j+1}\,+h_{n_{8}+2 {m_8}\, ; j-1} \,-h_{n_{8}+2 {m_8}\, \,; j+1}\right)
\nonumber \\,
\end{eqnarray}
where $a = a_1$ if $n_8$ is odd and  $a = a_2$ if $n_8$ is even.

Note that if the   $h_{n_{8}\,; j} \;$   satisfy the three term recurrence relation
\begin{equation}
\label{recur}
\sqrt{Y} \left(h_{n_{8}\,; j-1} \; - \, h_{n_{8}\,; j+1} \;\right)
=
\beta_{n_{8}} \;
\left[\frac{3}{20}+\frac{m^2}{2}+{n_8}^2+\frac{1}{9} \left(j-\sqrt{ {a}}\right)^2\right] \; h_{n_{8}\,; j},
\end{equation}
where the $\beta_{n_{8}}$ are also to be determined,
then
\begin{eqnarray}
\label{sc00rheq}
&0&=
\sum _{ {m_8}=-\infty}^{\infty } \frac{(-1)^{ {m_8}} \Gamma \left(\frac{5}{6}\right)^2 }{\Gamma \left(\frac{5}{6}- {m_8}\right) \Gamma \left( {\frac{5}{6}+m_8}\right)}
\; \left[\frac{3}{20}+\frac{m^2}{2}+ {\left({n_{8}-2 {m_8}}\right)}^2+\frac{1}{9} \left(j-\sqrt{ {a}}\right)^2\right]
\nonumber \\
&\times&
\left(\beta_{n_{8}-2 {m_8}}\,-\;\delta_{m_{8},0}\,\right)
\;
h_{n_{8}-2 {m_8}\,; j},
\end{eqnarray}
which is block diagonal in the temporal index $j$.

Except at the irregular  singular points $x\,=\,0\;\pm\pi$,
both
the massive complex scalar wave function $\psi$
and its quantum fluctuation $\delta \psi$ are expressible in the form given by

\begin{eqnarray}
\label{sc00FTsn3}
{\Psi}(\tau, x)&=&\,\,C_1\,{\Psi}_{\textrm{odd}}(\tau, x)\,+\,C_2\,{\Psi}_{\textrm{even}}(\tau, x)
\nonumber \\
{\Psi}_{\textrm{even}}(\tau, x) &=&
\sum_{n_{8}=-\infty,\,n_{8} \textrm{ even}}^{\infty} \sum_{j=-\infty}^{\infty} h_{n_{8}\,; j} \; \exp{ \left(\left(-\sqrt{a_2}\,+\, j\right)\, \tau \;\;+\;\dot{\imath}  \,n_{8}\, x\right)}
\nonumber \\
{\Psi}_{\textrm{odd}}(\tau, x) &=&
\sum_{n_{8}=-\infty,\,n_{8} \textrm{ odd}}^{\infty} \sum_{j=-\infty}^{\infty} h_{n_{8}\,; j}  \; \exp{ \left(\left(-\sqrt{a_1}\,+\, j\right)\, \tau \;\;+\;\dot{\imath}  \,n_{8}\, x\right)}
,
\end{eqnarray}
where the $h_{n_{8}\,; j}$
satisfy Eq.[\ref{sc00heq}].   Of course, the constants $C_1$ and $C_2$ are redundant and employed only for emphasis.
The scalar wave function  ${\Psi}(\tau, x)$ satisfies
\begin{eqnarray}
\label{sc00xeq}
0&=&
-{\Psi}^{(2,0)}(\tau,x)+\frac{1}{9} {\Psi}^{(0,2)}(\tau,x)
\nonumber \\
&+&
\left(
\frac{1}{2} m^2 + \frac{3 }{20}-
\,2\,
\sqrt{Y}\;
\sinh\left(\tau \right)\,\sqrt[6]{\csc^2 (x)}
\right)
\,{\Psi}(\tau,x)
.
\end{eqnarray}
The mass parameter $m$ is in general different for  $\Psi = \psi$ and  $\Psi = \delta \psi$.

A sequence of  approximate solutions to  Eq.[\ref{sc00heq}] may be defined as follows:
Let $n \in \mathbb{N}$, the natural numbers (excluding 0).
$n$ will be called the index of the approximation.
For each positive integer $n \in \mathbb{N} = 1, 2, \ldots$ we define an approximate solution
to  Eq.[\ref{sc00heq}]
in terms of
${\left(2 n +1\right)}^2$ mode coefficients
${{}^{(n)}}\-h_{n_{8}\,; j}$,
$-n \leq n_8, j \leq n$
that, in the limit $n \rightarrow \infty$,
will satisfy Eq.[\ref{sc00heq}] exactly.
\label{proc}
Eq.[\ref{sc00heq}] defines a  system of linear equations  for \emph{all} of the
$h_{n_{8}\,; j}$ mode coefficients.
In order to define a finite  approximate solution set,
we set all of the mode coefficients
$h_{n_{8}\,; j}=0$ for $-\infty \;< \;n_8, j <\;-n$ and $ n \;< \;n_8, j \;< \;\infty$.
This results in two homogeneous systems of linear equations  for the
${\left(2 n +1\right)}^2$ mode coefficients
${{}^{(n)}}\-h_{n_{8}\,; j} \,,\;-n \leq n_8, j \leq n$,
one system for the even $n_8$ modes and one system for the odd $n_8$ modes
(since
the even $n_8$ modes and odd $n_8$ modes evolve independently of each other).
To obtain non-trivial solutions to these two systems we
solve for the complex values of $a_1$ and $a_2$ that make the determinants of their respective
coefficient matrices vanish.
Two
dispersion relations thereby arise:
one that
relates $n$ and both $m$ and $k\,w$, the product of the magnitudes of the momentum wave vectors $(\vec{k}, \vec{w})$,
to the complex parameter $a_1$ for  the odd $n_8$ modes,
and
one that
relates $n$ and  $(m, \,k\,w)$
to the complex parameter $a_2$ for the even $n_8$ modes.

Recall that in this paper ``stable" means that
solutions $\Psi = \delta \psi$ of Eq.[\ref{sc00xeq}] possess a norm that is non-exponentially increasing as $\tau$ increases.
Stable modes exist when $-\textrm{Real}(\sqrt{a_1})+n$ and  $-\textrm{Real}(\sqrt{a_2})+n$ are both negative, or zero.
To see this
first
define the corresponding approximate solution ${{}^{(n)}\-\Psi}(\tau,x)$ to  Eq.[\ref{sc00xeq}] as
\begin{eqnarray}
\label{sc00FTsn}
{{}^{(n)}\-\Psi}_{\textrm{odd}}(\tau, x) &=&
\sum_{n_{8}=-n+evn(n),\,\textrm{ step 2}}^{n} \;\sum_{j=-n}^{n} \,{{}^{(n)}}\-h_{n_{8}\,; j}  \; e^{  \left[(\,-\sqrt{a_1}\,+\,j)\, \tau\;+\;\dot{\imath}  \,n_{8}\, x\right] }
\nonumber \\
{{}^{(n)}\-\Psi}_{\textrm{even}}(\tau, x) &=&
\sum_{n_{8}=-n+odd(n),\,\textrm{ step 2}}^{n} \;\sum_{j=-n}^{n} \,{{}^{(n)}}\-h_{n_{8}\,; j} \; e^{ \left[ (\,-\sqrt{a_2}\,+\,j)\, \tau\;+\;\dot{\imath}  \,n_{8}\, x\right] }
\nonumber \\
.
\end{eqnarray}
Here we have defined
two simple functions on $\mathbb{N}$, which for $\;n \in \mathbb{N}$, are given by
\begin{equation}
\begin{array}{cc}
evn(n) = \left\{\begin{array}{cc} 1&\textrm{ if } n \textrm{ is even}\\ 0&\textrm{ if } n \textrm{ is odd}\end{array}\right.
\;\;,\;\;
&
odd(n) = \left\{\begin{array}{cc}0&\textrm{ if } n \textrm{ is even}\\ 1&\textrm{ if } n \textrm{ is odd}\end{array}\right.
.
\end{array}
\end{equation}

One sees that, of all of the ${\left(2 n +1\right)}^2$ wave modes,
each of the form\\  ${{}^{(n)}}\-h_{n_{8}\,; j}\, e^{ \left[ (\,-\sqrt{a}\,+\,j)\, \tau\;+\;\dot{\imath}  \,n_{8}\, x\right] }$,
the $j = n$ waves increase at the greatest rate for given $a$.
If   $-\textrm{Real}(\sqrt{a_1})+n$ and  $-\textrm{Real}(\sqrt{a_2})+n$ are both negative, or zero,
then the $j = n$ waves have  non-increasing norms as time $\tau$ increases
(either because $w$ increases, $t$ increases or $k$ decreases),
so that the norm of ${{}^{(n)}\-\Psi}_{\textrm{odd}}(\tau, x) $
and ${{}^{(n)}\-\Psi}_{\textrm{even}}(\tau, x) $
also have  non-increasing norms as time $\tau$ increases
(here   $\Psi = \delta \psi$).
These stable modes
are associated to so-called
``stable zeros" $(a_1, a_2)$ of
the determinants of the coefficient matrices for
the odd $n_8$ labeled modes and
the even $n_8$ labeled modes, respectively.
The
``unstable zeros" $(a_1, a_2)$ of
the determinants of the coefficient matrices produce instabilities
driven by the momenta associated to the three extra time dimensions,
and do not yield physical solutions of the field equations.

In passing we remark that
the error of an approximation may be provisionally defined to be
\begin{eqnarray}
\label{sc00xerr}
\textrm{error}[\;{{{}^{(n)}\-\Psi}}(\tau,x)]
&=&
-{{{}^{(n)}\-\Psi}}^{(2,0)}(\tau,x)+\frac{1}{9} {{{}^{(n)}\-\Psi}}^{(0,2)}(\tau,x)
\nonumber \\
&+&
\left(
\frac{1}{2} m^2 + \frac{3 }{20}-
\,2\,
\sqrt{Y}\;  \sinh (\tau) \,\sqrt[6]{\csc^2 (x)}
\right)
\,{{{}^{(n)}\-\Psi}}(\tau,x)
.
\end{eqnarray}
However since
${{{}^{(n)}\-\Psi}}(\tau,x)$ contains only a finite number of Fourier modes it is reasonable to make a substitution
based on Eq.[\ref{CSC3approx}],
$\sqrt[6]{\csc ^2(x)}
\rightarrow
\frac{\sqrt[3]{2} \Gamma \left(\frac{2}{3}\right)}{\Gamma \left(\frac{5}{6}\right)^2}
{\sum _{{m8}=-\left[\frac{n}{2}\right]}^{\left[\frac{n}{2}\right]} \frac{e^{i (2 {m8}) x} \left((-1)^{{m8}} \Gamma \left(\frac{5}{6}\right)^2\right)}{\Gamma \left(\frac{5}{6}-{m8}\right) \Gamma \left({m8}+\frac{5}{6}\right)}}$,
in Eq.[\ref{sc00xerr}]
when computing the numerical error
($\left[\frac{n}{2}\right]$ gives the greatest integer less than or equal to $\frac{n}{2}$).
Since the error is expected to be large for small $n$ we defer error analysis.

\section{Coefficient matrices for the
${{}^{(n)}}\-h_{n_{8}\,; j}$ mode coefficients in Eq.[\ref{sc00heq}]} \label{cm}
Let
\begin{equation}\label{M}
M = \frac{1}{2} m^2 + \frac{3}{20},
\end{equation}
and recall that
$\sqrt{Y} = \frac{1}{2}k \, w \,
\sqrt[3]{2}\;
\frac{ \Gamma \! \left(\frac{2}{3}\right)}{\Gamma\!  \left(\frac{5}{6}\right)^2}$.
Fix $\;n \in \mathbb{N}$; 
we group the
${\left(2 n +1\right)}^2$=
${n\,\left(2 n +1\right)} $+
${\left( n +1\right)\,\left(2 n +1\right)} $ 
mode coefficients
${{}^{(n)}}\-h_{n_{8}\,; j} \,,\;-n \leq n_8, j \leq n$,
into two vectors 
$\overrightarrow{h}_{\textrm{even}}$ 
and
$\overrightarrow{h}_{\textrm{odd}}$
according to whether $n_8$ is even or odd. 
When $n$ is an odd positive integer then
$\overrightarrow{h}_{\textrm{odd}}$ has dimension 
${\left( n +1\right)\,\left(2 n +1\right)} = {\left( n +\,odd(n)\,\right)\,\left(2 n +1\right)}$ 
and
$\overrightarrow{h}_{\textrm{even}}$ has dimension 
${n\,\left(2 n +1\right)}  = {\left( n +\,evn(n)\,\right)\,\left(2 n +1\right)}$.
When $n$ is an even positive integer then
$\overrightarrow{h}_{\textrm{odd}}$ has dimension ${n\,\left(2 n +1\right)}  = {\left( n +\,odd(n)\,\right)\,\left(2 n +1\right)}$
and
$\overrightarrow{h}_{\textrm{even}}$ has dimension ${\left( n +1\right)\,\left(2 n +1\right)}  = {\left( n +\,evn(n)\,\right)\,\left(2 n +1\right)}$.
Note that $\overrightarrow{h}_{\textrm{odd}}$  always has an even number of components
and
that $\overrightarrow{h}_{\textrm{even}}$  always has an odd number of components.

Explicitly, the $i^{\textrm{th}}$ component of the vector $\overrightarrow{h}$ associated to ${{}^{(n)}}\-h_{n_{8}\,; j}$ has index
\begin{equation}\label{n8ji}
i\,=\frac{1}{2} (2 n+1)\left(n+\,{n_8} \, -\,evn(n)\,odd(n_8)\,-\,odd(n)\,evn(n_8)\,\right)+j+n+1
\end{equation}
so that
\begin{equation}\label{en8ji}
{h}_{\textrm{even}}\left(\frac{1}{2} (2 n+1)\left(n+\,{n_8} \, -\,odd(n)\,\right)+j+n+1\right) = 
{{}^{(n)}}\-h_{n_{8}\,; j},\;\textrm{where } n_8 \textrm{ is even }
\end{equation}
and
\begin{equation}\label{on8ji}
{h}_{\textrm{odd}}\left(\frac{1}{2} (2 n+1)\left(n+\,{n_8} \, -\,evn(n) \, \right)+j+n+1\right) =
{{}^{(n)}}\-h_{n_{8}\,; j},\;\textrm{where } n_8 \textrm{ is odd}.
\end{equation}
Here $h(i) = \left\{\overrightarrow{h}\right\}_{i} = i^{\textrm{th}} \textrm{ component of } \overrightarrow{h}$.

The inverse map
$(n;\,i, {parity}) \mapsto (n_8, j)$ is defined as follows:
Let \\
$
\mathbb{N} \ni parity = \left\{
\begin{array}{cc}
1 & \textrm{ if indices are extracted from } \overrightarrow{h}_{\textrm{odd}}\\
2 & \textrm{ if indices are extracted from } \overrightarrow{h}_{\textrm{even}}
\end{array}
\right.
$.

The inverse map 
$(n;\,i, \textrm{parity}) \mapsto (n_8, j)$ is given by
$j = \left(i -1 \right)\bmod (2 n+1)-n$
and 
$
n_8 = 2 \left[\frac{i-1}{2 n+1}\right]\,-\,n\,+\,odd(parity) evn(n)\,+\, evn(parity) odd(n)
$,
where
$\left[\frac{i-1}{2 n+1}\right]$
gives the greatest integer less than or equal to $\frac{i-1}{2 n+1}$.

Let $\mu$ denote the minimum value of $n_8$ for either
$\overrightarrow{h}_{\textrm{even}}$
or
$\overrightarrow{h}_{\textrm{odd}}$ (thus, $\mu$ is either equal to $-n$ or $-n+1$).
The components of these two vectors are ordered according to
$((n_8=\mu,j=-n),(n_8=\mu,j=-n+1),\ldots, (n_8=\mu,j=n),(n_8=\mu+2,j=-n),(n_8=\mu+2,j=-n+1),\ldots, (n_8=\mu+2,j=n),\ldots,
(n_8=-\mu,j=-n),(n_8=-\mu,j=-n+1),\ldots, (n_8=-\mu,j=n))$.

With respect the obvious canonical bases
each coefficient matrix is a square matrix 
%
%
and is equal to the sum $D + A$ of a diagonal matrix $D$, which carries the $a$ and $M$ dependence,
and an antisymmetric matrix   $\sqrt{Y}\, A$, where $\widetilde{A} = -A$ (the tilde denotes the transpose).
For a given $\;n \in \mathbb{N}$, the coefficient matrix 
${{}^{(n)}}\-D_{\textrm{even}} + {{}^{(n)}}\-A_{\textrm{even}}$ 
corresponding to
the even $n_8$ mode components
has dimensions  equal to
$\left[{\left( n +\,even(n)\right)\,\left(2 n +1\right)}\right] \times  \left[{\left( n +\,even(n)\right)\,\left(2 n +1\right)}\right]$,
and thus
always has an odd number of rows and columns.
The  coefficient matrix\\
$ {{}^{(n)}}\-D_{\textrm{odd}} + {{}^{(n)}}\-A_{\textrm{odd}}$  corresponding to
the  odd $n_8$ mode components
has dimensions  equal to
$\left[{\left( n +\,odd(n)\right)\,\left(2 n +1\right)}\right] \times  \left[{\left( n +\,odd(n)\right)\,\left(2 n +1\right)}\right]$,
and thus always has an even number of rows and columns.

The diagonal matrix ${{}^{(n)}}\-D$ has non-zero matrix elements
$
-\left(M\,+\,{n_8}^2\,+\,\frac{1}{9} \left(j-\sqrt{\text{a}}\right)^2\right)
$,
arranged (as described above) along the main diagonal in blocks, which are labelled by $n_8$, of the $2 n + 1$
possible $j$ values.

Because ${{}^{(n)}}\-A_{\textrm{even}}$ 
always has an odd number of rows and columns it possesses a vanishing determinant,
$\det{\left({{}^{(n)}}\-A_{\textrm{even}}\right)} \equiv 0$.
The  antisymmetric ${{}^{(n)}}\-A_{\textrm{odd}}$  corresponding to
the  odd $n_8$ mode components always has an even number of rows and columns, and hence a 
possibly non-vanishing determinant. However for this case the rows  \\
$
{{}^{(n)}}\-A_{\textrm{odd}}(i),\,i= 1, 2, \ldots,
\left[ n +odd(n)\right]\times \left(2 n +1\right) 
$ 
of ${{}^{(n)}}\-A_{\textrm{odd}}$
are not linearly independent and satisfy
\begin{equation}
{{}^{(n)}}\-A_{\textrm{odd}}(1)+ {{}^{(n)}}\-A_{\textrm{odd}}(3)+\cdots+ {{}^{(n)}}\-A_{\textrm{odd}}(2 n +1)
=
\stackrel{{\left( n +odd(n)\right)\,\left(2 n +1\right)}}{\overbrace{(0, 0, \ldots, 0)}}.
\end{equation}
Hence for this case $A_{\textrm{odd}}$ also has a vanishing determinant,
$\det{\left({{}^{(n)}}\-A_{\textrm{odd}}\right)} = 0$.

The antisymmetric matrix ${{}^{(n)}}\-A$  is related to
an antisymmetric Toeplitz matrix\\
${{}^{(n)}}\-T = -\widetilde{{{}^{(n)}}\-T\phantom{W}}$ whose non-zero upper triangular matrix elements 
are given by\\
${{}^{(n)}}\-T_{i\,,i+\,\ell\,\left(2 n + 1\right)+1} = -\gamma_\ell,\;\ell = 0, 1, \dots$,
and\\
${{}^{(n)}}\-T_{i\,,i+\,\ell\,\left(2 n + 1\right)-1} = \gamma_\ell,\;\ell = 1, \dots$,
where the $\gamma_\ell$ are defined in Eq.[\ref{gam}]
and
$i+\,\ell\,\left(2 n + 1\right)+1 \leq  \left[{\left( n +\,odd(n)\right)\,\left(2 n +1\right)}\right]$ for
${{}^{(n)}}\-T$ corresponding to the ${{}^{(n)}}\-A_{\textrm{odd}}$ associated to the odd $n_8$ mode components
and\\
$i+\,\ell\,\left(2 n + 1\right)+1 \leq  \left[{\left( n +\,evn(n)\right)\,\left(2 n +1\right)}\right]$ for
${{}^{(n)}}\-T$ corresponding to the ${{}^{(n)}}\-A_{\textrm{even}}$ associated to  the  even $n_8$ mode components.
The remaining matrix elements of ${{}^{(n)}}\-T$ are either dictated by antisymmetry, or are zero.

The  matrix elements of ${{}^{(n)}}\-A$ equal the matrix elements of ${{}^{(n)}}\-T$
except that they are
punctuated by additional zeros along the super diagonals of ${{}^{(n)}}\-T$.
${{}^{(n)}}\-A_{\textrm{even}},\;n \geq 2,$ has 
$N_{\textrm{additional zeros}} = N_{AZ} = 2\left(\left[n + evn(n)\right]^2 - 1\right)$ replacements
of non-zero matrix elements of ${{}^{(n)}}\-T$ by zero along its super diagonals.
${{}^{(n)}}\-A_{\textrm{odd}},\;n \geq 1,$ has $N_{AZ} = 2\left(\left[n + odd(n)\right]^2 - 1\right)$ zero replacements
of non-zero matrix elements of ${{}^{(n)}}\-T$ by zero along its super diagonals.

The specification of the row and column indices $(r, c)$ of a particular matrix element\\  
${{}^{(n)}}\-A_{r\,c}\;$   
of   $\;\;{{}^{(n)}}\-A$ that is zero, when  ${{}^{(n)}}\-T_{r\,c}$ is not zero, is
somewhat involved, but may be described in stages.
Let the dimensions of   ${{}^{(n)}}\-A$  be $L \times  L$.
Define a mapping of the two-dimensional array
${{}^{(n)}}\-A_{r\,c},\;r, c= 1, \ldots, L$   
to a one-dimensional array \\
${{}^{(n)}}\-B_{i},\;i= 1, \ldots, L^2$ by setting
${{}^{(n)}}\-B_{i} = {{}^{(n)}}\-B_{(r-1)L+c} = {{}^{(n)}}\-A_{r\,c}$.
Formally, $i = (r-1)L+c$.
The inverse map
$(n;\,i, \textrm{parity}) \mapsto (r, c)$ is given by
$r = \left(i -1 \right)\bmod (L) + 1$
and
$c = \left[\frac{i-1}{L}\right]\,+\, 1$,
where
$L=\left(2 n+1\right) \left(n + evn(parity)\,evn(n)+odd(parity) \,  odd(n)\right)$.

Assume that   
the row and column indices $\left.\left\{(r, c)_{j}\right\}\right]_{j=1}^{N_{AZ}}$ of the
$N_{AZ}$ matrix elements whose non-zero 
${{}^{(n)}}\-T_{r_j\;c_j}$ values are to be
replaced by zero are known, and that their corresponding 
$\left.\left\{i_{j}\right\}\right]_{j=1}^{N_{AZ}}$ values have been computed;
sort the  $\left\{i_{j}\right\}$ values in ascending order.
Define the relative displacement coordinates $\left.\left\{I^{\textrm{RDC}}_j\right\}\right]_{j=1}^{N_{AZ}}$ 
by
$I^{\textrm{RDC}}_1 = i_1$,
$I^{\textrm{RDC}}_2 = i_2-i_1,\;\ldots$,
$I^{\textrm{RDC}}_j = i_j-i_{j-1},\;j=2,\ldots,N_{AZ}$.
The inverse map is simply
$i_j = \sum_{h=1}^{j}{I^{\textrm{RDC}}_h}$.

We find that the $\left.\left\{I^{\textrm{RDC}}_j\right\}\right]_{j=1}^{N_{AZ}}$  are given by  two sequences,
one for the  odd $n_8$ mode components and one for the  even $n_8$ mode components.
Let $N_j = 2 n + 1$ denote the number of $j$-values in the $n^{\textrm{th}}$-order approximation.
Let $\left\{N_j\right\}[h]$ denote the finite sequence 
$\left\{N_j\right\}[h]
=
\stackrel{h}{\overbrace{\left\{N_j, N_j, \ldots, N_j\right\}}}
$
with $h$ elements, and let
$\{S\}[h]$  denote a finite sequence repeated $h$ times.

For the case of the odd $n_8$ mode components let \\$N = n - evn(n)$
and\\
$\alpha =
\frac{1}{2} \left(2 n \left(2 n \left(2 n-(-1)^n+1\right)-1\right)+(-1)^n+1\right)$
; the  sequence 
$\left.\left\{I^{\textrm{RDC}}_j\right\}\right]_{j=1}^{N_{AZ}}$ for the odd $n_8$ mode components is
\begin{eqnarray}
\label{RDCo}
\left.\left\{I^{\textrm{RDC}}_{\textrm{odd}\;j}\right\}\right]_{j=1}^{N_{AZ}}
&=&
\left\{
\left\{N_j\right\}[N], \alpha+N_j,
\left\{\left\{N_j\right\}[N], 1+4 n, \left\{N_j\right\}[N], \alpha
\right\}[N-1],
\right.
\nonumber \\
&&
\left.
\left\{N_j\right\}[N], \alpha+N_j,\left\{N_j\right\}[N]
\right\}
.
\end{eqnarray}

For the case of the  even $n_8$ mode components let \\$N = n - odd(n)$
and\\
$\beta =
\frac{1}{2} \left(2 n \left(2 n \left(2 n+(-1)^n+1\right)-1\right)-(-1)^n+1\right)$
; the  sequence
$\left.\left\{I^{\textrm{RDC}}_j\right\}\right]_{j=1}^{N_{AZ}}$ for the even $n_8$ mode components is
\begin{eqnarray}
\label{RDCe}
\left.\left\{I^{\textrm{RDC}}_{\textrm{even}\;j}\right\}\right]_{j=1}^{N_{AZ}}
&=&
\left\{
\left\{N_j\right\}[N], \beta+N_j,
\left\{\left\{N_j\right\}[N], 1+4 n, \left\{N_j\right\}[N], \beta
\right\}[N-1],
\right.
\nonumber \\
&&
\left.
\left\{N_j\right\}[N], \beta+N_j,\left\{N_j\right\}[N]
\right\}
.
\end{eqnarray}

Only the first $N_{AZ}$ sequence elements are to be employed when evaluating the
the relative displacement coordinates from the previous two sequence formulas.

\section{Theorem: Existence of stable solutions}
%
We record the
\begin{theorem}
\label{tj}
Let $n \in \mathbb{N}$ be the index of an approximation.
Define an approximate solution as described above
to  Eq.[\ref{sc00heq}]
in terms of
${\left(2 n +1\right)}^2$ mode coefficients
${{}^{(n)}}\-h_{n_{8}\,; j}$,
$-n \leq n_8, j \leq n$.

Let
${{}^{(n)}\-D}_{n_8\,\textrm{ odd}}(a_1, Y, M)$
denote
the determinant of the coefficient matrix of the ${{}^{(n)}}\-h_{n_{8}\,; j}$
for
the odd $n_8$ labeled modes
and
${{}^{(n)}\-D}_{n_8\,\textrm{ even}}(a_2, Y, M)$
denote the determinant of the coefficient matrix of the ${{}^{(n)}}\-h_{n_{8}\,; j}$
for
the even $n_8$ labeled modes.

Let  $n_{a_{1}}$ denote the number of stable zeros of ${{}^{(n)}\-D}_{n_8\,\textrm{ odd}}(a_1, Y, M)$
and
$n_{a_{2}}$ denote the number of stable zeros of  ${{}^{(n)}\-D}_{n_8\,\textrm{ even}}(a_2, Y, M)$.

Then

\begin{enumerate}
\label{thm}
\item {${{}^{(n)}\-D}_{n_8\,\textrm{ odd}}(a_1, Y, M)$ is a trivariate polynomial in  $(a_1,\,Y,\,M)$ with rational coefficients};\label{thm1}
\item {${{}^{(n)}\-D}_{n_8\,\textrm{ even}}(a_2, Y, M)$ is a trivariate polynomial in  $(a_2,\,Y,\,M)$ with rational coefficients};\label{thm2}
\item {For each positive integer $ n \in \mathbb{N} = 1, 2, \ldots$,
stable zeros  $(a_1, a_2)$ of the
the determinants of the coefficient matrices
${{}^{(n)}\-D}_{n_8\,\textrm{ odd}}(a_1, Y, M)$
and
${{}^{(n)}\-D}_{n_8\,\textrm{ even}}(a_2, Y, M)$
exist that satisfy
$-\textrm{Real}(\sqrt{a_1})+n\,\leq\,0$ and  $-\textrm{Real}(\sqrt{a_2})+n\,\leq\,0$
};
\item {if $n$ is an odd positive integer then}
\begin{enumerate}
\item ${{}^{(n)}\-D}_{n_8\,\textrm{ odd}}(a_1, Y, M)$ is of degree (n+1)(2 n+1) in $a_1$ and $M$, and of degree n(n+1) in Y
\item ${{}^{(n)}\-D}_{n_8\,\textrm{ even}}(a_2, Y, M)$ is of degree n(2 n+1) in $a_2$ and $M$, and of degree $n^2$ in Y
\begin{itemize}
\item If $\;Y \,= \,1$ then
\begin{enumerate}
\item $n_{a_{1}}$ =  2 n+2
\item $n_{a_{2}}$ =  2 n
\end{enumerate}
\item If $\;Y \,\neq \,1$ then
\begin{enumerate}
\item $n_{a_{1}} \geq  2 n+2$
\item $n_{a_{2}} \geq  2 n$
\end{enumerate}
\end{itemize}
\end{enumerate}
\item {if $n$ is an even positive integer then}
\begin{enumerate}
\item ${{}^{(n)}\-D}_{n_8\,\textrm{ odd}}(a_1, Y, M)$ is of degree n(2 n+1) in $a_1$ and $M$, and of degree $n^2$ in Y
\item ${{}^{(n)}\-D}_{n_8\,\textrm{ even}}(a_2, Y, M)$ is of degree (n+1)(2 n+1) in $a_2$ and $M$, and of degree n(n+1) in Y
\begin{itemize}
\item If $\;Y \,= \,1$ then
\begin{enumerate}
\item $n_{a_{1}}$ = 2 n
\item $n_{a_{2}}$ = 2 n+2
\end{enumerate}
\item If $\;Y \,\neq \,1$ then
\begin{enumerate}
\item $n_{a_{1}} \geq 2 n$
\item $n_{a_{2}} \geq 2 n+2$
\end{enumerate}
\end{itemize}
\end{enumerate}
\end{enumerate}

This relationship holds in the limit $n \rightarrow \infty $
yielding
exact stable solutions to  Eq.[\ref{sc00heq}], and therefore to   Eq.[\ref{sc00xeq}].
\end{theorem}

Partial Proof of  Theorem[\ref{tj}]:
For $n = 1, \ldots, 9$ the proof has been accomplished by direct calculation
and is summarized in Table[\ref{teeble}].

In general, parts [\ref{thm1}] and [\ref{thm2}]
may be proven by employing Theorem 13.7.3. of \cite{opac-b1094615}, which gives 
the  expansion for the determinant $|D + B|$
of the sum of
two $n \times n$ matrices $D$ and $B$ for the special case where one of the matrices is diagonal.
We quote Theorem 13.7.3. \cite{opac-b1094615}:
Let $B$ represent an  $n \times n$  matrix, and let $D$ represent an  $n \times n$ 
diagonal matrix whose diagonal elements are $d_1, \ldots , d_n$. Then,

\begin{equation}\label{thmH}
|D + B|
=
\sum_{\{i_1, i_2, \ldots,i_r\}}d_{i_1}\cdot d_{i_2}\cdots \;\cdot d_{i_r}|B^{\{i_1, i_2, \ldots,i_r\}}|.
\end{equation}
where $\{i_1, i_2, \ldots,i_r\}$ is a subset of the first n positive integers $1, \ldots , n$ (and the
summation is over all $2^n$ such subsets) and where $B^{\{i_1, i_2, \ldots,i_r\}}$ is the 
$\left(n - r\right) \times \left(n - r\right)$
principal submatrix of $B$ obtained by striking out the ${i_1, i_2, \ldots,i_r}$th rows and columns.
[The term in the sum [\ref{thmH}] corresponding to the empty set is to be interpreted as  $|B|$, and
the term corresponding to the set $\{1, 2, \ldots , n\}$ is to be interpreted as  
$|D | =d_{1}\cdot d_{2}\cdots \;\cdot d_{n}$.]

In our case the matrix elements of $D$ and $B$ are defined in Section [\ref{cm}].
$B = \,\sqrt{Y}\,A$ is antisymmetric. Note that all principal submatrices  $B^{\{i_1, i_2, \ldots,i_r\}}$ 
of the antisymmetric matrix  $B = \,\sqrt{Y}\,A$ are also antisymmetric.

The general proof of this theorem is a work in progress.
Thus, for all other cases than those explicitly noted, the Theorem devolves to a Conjecture.

Note that according to
Theorem[\ref{tj}], for given $n$, the degrees of the $\{a_1, a_2\}$ equal the numbers of
independent
symmetric and anti-symmetric matrix elements of a $(2 n +1) \, \times (2 n + 1)$
real matrix. The sum of the degrees of the $(a_1, a_2)$ equals ${(2 n + 1)}^2$.

Table [\ref{teeble}] summarizes the degrees of the  polynomials
${{}^{(n)}\-D}_{n_8\,\textrm{ odd}}(a_1, Y, M)$ and ${{}^{(n)}\-D}_{n_8\,\textrm{ even}}(a_2, Y, M)$
found by direct calculation
for $n = 1, \ldots, 9$.
The degree of
$
M = \frac{1}{2} m^2 + \frac{3}{20}
$
is found to always equal the degree of $a_1$ for
${{}^{(n)}\-D}_{n_8\,\textrm{ odd}}(a_1, Y, M)$
and
always equal the degree of $a_2$ for
$ {{}^{(n)}\-D}_{n_8\,\textrm{ even}}(a_2, Y, M)$,
and therefore is not displayed in this table.

\begin{equation}\label{teeble}
\begin{tabular}{|l||*{4}{c|}||c|c|}\hline
&\multicolumn{2}{|c|}{ ${{}^{(n)}\!D}_{n_8\,\textrm{ odd}}(a_1, Y, M)$ }
&\multicolumn{2}{|c|}{ ${{}^{(n)}\!D}_{n_8\,\textrm{ even}}(a_2, Y, M)$ }
&\multicolumn{2}{|||c|}{ \# STABLE }
\\\hline
Polynomial& ${a_1}^{\alpha}$ &  ${Y}^{\beta}$ & ${a_2}^{\gamma}$ &  ${Y}^{\delta}$& \multicolumn{2}{|||c|}{for Y=1 } \\ \hline
\backslashbox{n}{Degree}& $\alpha$ &  $\beta$ &  $\gamma$ &  $\delta$  &  $n_{a_{1}}$ &  $n_{a_{2}}$ \\ \hline\hline
1 &6&2&3&1&4&2\\\hline
2 &10&4&15&6&4&6\\\hline
3 &28&12&21&9&8&6\\\hline
4 &36&16&45&20&8&10\\\hline
5 &66&30&55&25&12&10\\\hline
6 &78&36&91&42&12&14\\\hline
7 &120&56&105&49&16&14\\\hline
8 &136&64&153&72&16&18\\\hline
9 &190&90&171&81&20&18\\\hline
& \multicolumn{6}{|c|}{ CONJECTURE }\\\hline
odd n&(n+1) (2 n+1)&n (n+1)&n (2 n+1)&$n^2$&2 n+2&2 n\\\hline
even n&n (2 n+1)&$n^2$&(n+1) (2 n+1)&n (n+1)&2 n&2 n+2\\\hline
\end{tabular}
\end{equation}

The significance of this theorem is that each pair of ``stable" zeros  $(a_1, a_2)$ defines a pair of stable propagation modes for
a quantum fluctuation $\delta \psi$.
The numbers of these modes increases with $n$
as $n \rightarrow \infty$
yielding a possibly complete set of decaying quasi-normal modes
that one may employ in cosmological perturbation calculations.
Quantitative calculations of the quantum fluctuations that occur during inflation
are deferred until we have an analytical approximation for
a complete set of decaying quasi-normal mode functions for the massive complex scalar field.

\subsection{Examples for $Y = 1, \,m = 0$}
The  results of an example for $n = 6$,
$m = 0$
and
$\sqrt{Y} = \frac{1}{2}k \, w \,
\sqrt[3]{2}\;
\frac{ \Gamma \! \left(\frac{2}{3}\right)}{\Gamma\!  \left(\frac{5}{6}\right)^2} = 1$
are displayed in Figures [2] and [3], which
present pairs of plots ($\tau$ dependence, for fixed $x$)  and ($x$ dependence, for fixed $\tau$)
of the stable scalar wave functions ${{{}^{(6)}\!\psi}}(\tau, x)$ of Eq.[\ref{sc00FTsn}].
For $n = 6$ there are $169 = (2 n + 1)^2$ mode coefficients,
78 with odd $n_8$ and 91 with even $n_8$.
Roughly 352 digits of precision in the solved-for values of $a_1$ and $a_2$ are required to make the
determinants of the coefficient matrices
vanish to double precision accuracy.
The relevant values (rounded to 9 digits) of $a_1$ and $a_2$ for each plot are listed in the yellow box associated to the plot; space limitations
do not permit their display in a table
(if one is reading the pdf version of the manuscript then zooming-in may be required, unless you have remarkable eyesight).
Red curves correspond to the real part of the solutions and blue curves
and correspond to the imaginary part.

The  results of an example for $n = 7$,
$m = 0$
and
$\sqrt{Y} = \frac{1}{2}k \, w \,
\sqrt[3]{2}\;
\frac{ \Gamma \! \left(\frac{2}{3}\right)}{\Gamma\!  \left(\frac{5}{6}\right)^2} = 1$
are displayed in Figures [4] and [5], which also
present pairs of plots ($\tau$ dependence, for fixed $x$)  and ($x$ dependence, for fixed $\tau$)
of the stable scalar wave functions ${{{}^{(7)}\!\psi}}(\tau, x)$ of Eq.[\ref{sc00FTsn}].
For $n = 7$ there are $225 = (2 n + 1)^2$ mode coefficients,
120 with odd $n_8$ and 105 with even $n_8$.
In this case,  about 400 digits of precision in the solved-for values of $a_1$ and $a_2$ are required to make the
determinants of the coefficient matrices
vanish to double precision accuracy.
The relevant  values of $a_1$ and $a_2$ for a plot are again listed in the yellow box
(zooming-in may also be required, unless you have remarkable eyesight).
Red curves correspond to the real part of the solutions and blue curves
and correspond to the imaginary part.

The  results of an example for $n = 8$,
$m = 0$
and
$\sqrt{Y} = \frac{1}{2}k \, w \,
\sqrt[3]{2}\;
\frac{ \Gamma \! \left(\frac{2}{3}\right)}{\Gamma\!  \left(\frac{5}{6}\right)^2} = 1$
are displayed in Figures [6] and [7], which
present pairs of plots ($\tau$ dependence, for fixed $x$)  and ($x$ dependence, for fixed $\tau$)
of the stable scalar wave functions ${{{}^{(8)}\!\psi}}(\tau, x)$ of Eq.[\ref{sc00FTsn}].
For $n = 8$ there are $361 = (2 n + 1)^2$ mode coefficients,
190 with odd $n_8$ and 171 with even $n_8$.
In this case,  about 400 digits of precision in the solved-for values of $a_1$ and $a_2$ are required to make the
determinants of the coefficient matrices
vanish to double precision accuracy.
The relevant  values of $a_1$ and $a_2$ for a plot are again listed in the yellow box
(zooming-in may also be required, unless you have remarkable eyesight).
Red curves correspond to the real part of the solutions and blue curves
and correspond to the imaginary part.

Qualitatively, corresponding $n = 6$, $n = 7$ and $n = 8$ plots agree,
although there are quantitative differences due to the low orders of approximation.
The overall number of modes increases by four for each increment in $n$.
In the limit $n \rightarrow \infty$ it is reasonable to hope
that this procedure generates a ``complete" set of stable modes,
in the sense that any \textbf{stable} wavefunction can be expressed as a
linear combination of these modes.

We arrive at a  picture of stable “quasi-normal modes” that decay during inflation.
According to  Theorem[\ref{tj}]   this model possesses an infinite number of stable “quasi-normal modes”,
such that
the momenta associated with the extra time dimensions do not create instability.

\section{Class $E_1$ solutions of the Einstein equations}
\label{sec:C1}
The  field equations admit a second class of solutions
that is periodic in  $x^8$ and
parameterized by
$\xi$,  $-\frac{1}{\sqrt{ 5 } } < \xi < \frac{1}{\sqrt{ 5 } }$,
and which has a time dependent $\ell$.
In this case the scale factors are
\begin{eqnarray}
\label{fyxx1}
a
&=&
\nonumber \\
&&
a_{1}
\left[\cosh \left(x^4 \,\sqrt{\frac{\Lambda }{3} }\right) \right]^{ \frac{1}{6} \left(1 + 5 \xi\right)}
\nonumber \\
&\times&
\left[\tan ^2 \left(\frac{1}{2} \sqrt{\frac{5}{3}} \sqrt{\Lambda } x^8\right) \right]^{\pm \sqrt{\frac{1}{30}} \sqrt{1-5 \xi ^2}}
{
\left[\sin ^2 \left(\sqrt{\frac{5}{3}} \sqrt{\Lambda } x^8\right)\right]^{  \frac{1}{12}  \left(1 - \xi\right)}}
\nonumber \\
b
&=&
\nonumber \\
&& b_{1}
\left[\cosh \left(x^4 \,\sqrt{\frac{\Lambda }{3} }\right) \right]^{ \frac{1}{6}  \left(1 - 5 \xi\right)}
\nonumber \\
&\times&
\left[\tan ^2 \left(\frac{1}{2} \sqrt{\frac{5}{3}} \sqrt{\Lambda } x^8\right) \right]^{\mp \sqrt{\frac{1}{30}} \sqrt{1-5 \xi ^2}}
{
\left[\sin ^2 \left(\sqrt{\frac{5}{3}} \sqrt{\Lambda } x^8\right)\right]^{ \frac{1}{12}  \left(1 + \xi\right)}}
,
\nonumber \\
\end{eqnarray}
where $a_1 $ and $b_1 $ are constants.
Periodicity in  $x^8$ requires that $\Lambda$ be quantized,
\begin{equation}
\label{L1}
\Lambda\,=\,\Lambda_1\,=\,\frac{3}{5} \,=\,\frac{6}{5} \,\Lambda_0
\end{equation}
We also find that
\begin{eqnarray}
\label{fyyxx1}
\varphi
&=&
\frac{\sqrt{1-5\, \xi ^2} \ln \left[\cosh ^5\left(\frac{\sqrt{\Lambda } x^4}{\sqrt{3}}\right) \csc^2 \left(\sqrt{\frac{5}{3}} \sqrt{\Lambda } x^8\right)\right]}{2\sqrt{30}}
\mp \, \xi  \ln \left[\tan ^2 \left(\frac{1}{2} \sqrt{\frac{5}{3}} \sqrt{\Lambda } x^8\right)\right]
\nonumber \\
\ell_1 &=& \mp\,\frac{5 \sqrt{\Lambda } \xi  \tanh \left(\frac{\sqrt{\Lambda } x^4}{\sqrt{3}}\right)}{3 \sqrt{3}}
=
\pm\,
5\,\xi\,\sqrt{\frac{2}{3}} \; \ell_0 \,  \tanh \left(x^4 \,\sqrt{\frac{\Lambda }{3} }\right).
\end{eqnarray}

To exclude   collapsing universe solutions, which correspond to
$\frac{\partial}{\partial x^4} \ln{\left( a \right)} < 0$,
and solutions for  universes with multiple macroscopic times, which  correspond to
$\frac{\partial}{\partial x^4} \ln{\left( b \right)} \ge 0$,
from the set
of physical solutions
we restrict  $\xi$ to the interval  $ \frac{1}{{ 5 } } < \xi < \frac{1}{\sqrt{ 5 } }$.

For this case the volume element $d \Omega =  d \tau\, \cosh \left(x^4 \,\sqrt{\frac{\Lambda }{3} }\right)  \left| {\sin  \left(\, x^8 \,\sqrt{\frac{5}{3}\,\Lambda } \,\right)}\right|$,
is  independent of both $\xi$ and the choice of $\pm$ signs.
The temperature history during inflation is non-trivial in this case.
This and the possible stability of quantum fluctuations are of interest, but  not investigated further in this paper.

\section{Conclusion}
\label{sec:work}

We have presented a model of the very early universe that possesses  equal numbers of space
and time dimensions.
In this model, after ``inflation" the observable physical macroscopic world appears to a classical observer to
be a homogeneous, isotropic universe with three space dimensions and one time dimension.
Notwithstanding the often-expressed concern that the momenta associated to  extra time dimensions
source destabilizing quantum fluctuations,
we have shown that
the physical solutions $\delta \psi$ to Eq.[\ref{sc00xeq}] that propagate on the class  $E_0$ solution to the
coupled Einstein-inflaton equations possess decaying quasi-normal modes,
and have no instabilities that
are sourced by the momenta $\vec{w}$ associated to the extra time dimensions.

This model and the class  $E_0$ solution to the
coupled Einstein-inflaton field equations provide simple answers to at least three
of the important questions about the early universe,
namely,
how to the define the inflaton potential \cite{RevModPhys.69.373},
how to the define the inflaton mass
and how to resolve the issue of reheating  \cite{PhysRevD.56.3258}.
At the classical level the answers to the first two questions are zero inflaton potential and zero inflaton mass.
Computable quantum fluctuations may shift the classical $V_{\textrm{inflaton}} = 0$ value
and the classical $m_{\textrm{inflaton}} = 0$ value.
How these values behave under renormalization are important questions, but beyond the scope
of the present paper.
The answer to the last question is  that there is no obvious need for reheating, since  inflation/deflation is isothermal
in this model for the class  $E_0$ solution (the ground state solution to the Einstein-inflaton field equations).

We have also proved that a well known so-called ``single time" theorem  \cite{:/content/aip/journal/jmp/40/2/10.1063/1.532695}
\textbf{does not apply}
to our model.

Do closed timelike curves exist in this model?
If we assume that an ``arrow of time" exists for each timelike dimension then
this model does not admit unphysical closed timelike curves.
The reason is simply
that the tangent vector to any such curve is not everywhere future-directed, if it is timelike.
Let $x^{\alpha}=x^{\alpha}(\xi),\;\;0 \leq \xi \leq 1$,
with  $x^{\alpha}(0)=x^{\alpha}(1)$,
parameterize a timelike closed curve in
$\mathbb{X}_{4,4}$;
then
$\frac{\partial}{\partial\,\xi}x^{\alpha}(\xi)$
for $\;0 \leq \xi \leq 1$
is not everywhere either  future-directed timelike, null, or spacelike.
For example,
a closed timelike curve within the $(x^4, x^5, x^6, x^7)$ 4-manifold  that traverses the $x^4$ dimension
is not physical classically because such a curve must move both forward and backward in physical time $x^4$.

If we do not assume that an ``arrow of time" exists for each extra timelike dimension then, if
deflation for the
scale factor associated to the extra time dimensions
proceeded for sufficiently long $\Delta\,x^4$ interval(s),
then observationally the effects of closed timelike curves
might be described as a difficulty in assigning classical spacetime coordinates to  certain events.
A discussion of this is beyond the scope of this paper.

In this model as presently formulated, inflation also goes on forever.
This ``feature" obviously must be eliminated from the model.
This model must be generalized in a way that leads to inflaton decay and also
terminates inflation.
Possible generalizations include:  [1] modifing the inflaton Lagrangian to include a self-coupling; and
[2] coupling the inflaton to a Higgs doublet
(before the Higgs develops a non-zero vacuum expectation value, in order to keep the inflaton massless) through a Yukawa interaction, while also  generalizing the total Lagrangian in the model to  incorporate
Standard Model quarks, leptons and gauge bosons.

\subsection{The scale factors vanish on a set of measure zero}
\label{fluctmz}

Typically and approximately, inflation scenarios inflate a scale of the size of one billionth the present radius of a proton to
the size of the present radius of a marble or a grapefruit in about  $10^{-32}$   seconds.
In virtue of the Heisenberg Uncertainty Principle,
and because the comoving  $(x^1, x^2, x^3)$ dimensions have undergone inflation
while the $x^8$ dimension has not,
present epoch quantum fields that are functions of $(x^1, x^2, x^3, x^4, x^8)$
are expected to almost uniformly sample the region of the $x^8$ dimension that they occupy.
The spatial $x^8$ average of functions of $x^8$ are expected to appear in   effective four dimensional spacetime theories.
The fact that the scale factors vanish on a set of
measure zero may be handled in a straightforward manner by employing spatial $x^8$ averages
in physical calculations.

\appendix

\section{Solution of the diagonal contribution to the mode equations (neglecting mode coupling)}\label{sec:esss4}

For both the
$\zeta = 3/10$ case
and the second case with
$\zeta = 0$, one may  solve the diagonal mode equations
obtained from  Eq.[\ref{mainFnEQ}] and  Eq.[\ref{escact02yn}]
using Mathieu functions.
However complex manipulations involving such mode functions is limited by the state of Mathieu function science.
Instead, for both cases,
to  solve the uncoupled mode function equations we make use of the following identity,
which is easily verified using the recursion relations for the Bessel functions.
First, let us  define several quantities.

Let ${m_{\textrm{eff}}}^2>0, B>0, \nu, \alpha$ and $\beta$ be given  constants, and $n \in \mathbb{Z}$; let
$J_n = J_{{n}}\left(3\,k\,  \sqrt{2 \,{B}} \; \exp \left(-\frac{t}{6}\right)\right)$,
$K_{\mu} = K_{{\mu}}\left( 3\,w\, \sqrt{2 \,{B}} \; \exp \left( +\frac{t}{6}\right)\right)$
and
$I_{\mu} = I_{{\mu}}\left( 3\,w\, \sqrt{2\, {B}} \; \exp \left( +\frac{t}{6}\right)\right)$ .
Then  using the recursion relations for  the Bessel functions \cite{MR0010746} it is straightforward to show that

\begin{eqnarray}
\label{JKidentity}
0&=&
\frac{\partial ^2}{\partial t^2}
\left[
\left(\alpha  K_{{n}+\dot{\imath}\,\nu }\,+\beta  I_{{n}+\dot{\imath}\,\nu }\,\right)\,
J_{{n}}\,
\right]
+
\nonumber \\
&&
\left[{{m_{\textrm{eff}}}^2}- {B} \frac{1}{2}\left(w^2\, e^{t/3} \,-k^2 \,e^{-t/3}  \,\right)\right]
\left[
\left(\alpha  K_{{n}+\dot{\imath}\,\nu }\,+\beta  I_{{n}+\dot{\imath}\,\nu }\,\right)\,
J_{{n}}\,
\right]\,
\nonumber \\
&-&
\left[ {m_{\textrm{eff}}}^2 +\frac{1}{9} \left(n+\frac{\dot{\imath}\,\nu }{2}\right)^2\right]
\left[
\left(\alpha  K_{{n}+\dot{\imath}\,\nu }\,+\beta  I_{{n}+\dot{\imath}\,\nu }\,\right)\,
J_{{n}}\,
\right]\,
\nonumber \\
&+&
\left.
\frac{1}{2} {\,k \,w \,B\,}
\left(-\alpha  K_{{n-1}+\dot{\imath}\,\nu }\,+\beta  I_{{n-1}+\dot{\imath}\,\nu }\,\right)
\,J_{{n}-1}\,
\right.
\nonumber \\
&+&
\frac{1}{2} {\,k \,w \,B\,}
\left(\phantom{-}\alpha  K_{{n+1}+\dot{\imath}\,\nu }\,-\beta  I_{{n+1}+\dot{\imath}\,\nu }\,\right)
\,J_{{n}+1}\, .
\end{eqnarray}

Employing this identity to solve the uncoupled mode function equations
has its roots in a closely related technique due to Dougall
\cite{dougall1915solution},
Section[15], pages 191-193,
with appropriate modifications that account for the
asymmetric ``potential"
$\frac{1}{2}\left( \, w^2\,e^{t/3} \,- \,k^2\, e^{-t/3}\right)$
in this problem.
Dougall gives the solution of Mathieu's modified differential equation as a series of products of Bessel functions.

The general solution to Eq.[\ref{modeD1}],
\begin{displaymath}
0 =
\ddot{Y}_{m_8}(t)+
\left[
{m_{\textrm{eff}}}^2\,-\,B\,\frac{1}{2}\left( \, w^2\,e^{t/3} \,- \,k^2\, e^{-t/3}\right)\,
\right]{Y}_{m_8}(t),
\end{displaymath}
may be obtained by
introducing
coefficient sets $\{c_n,{d_n}\}\,,\;n= -\infty,\ldots, \infty$
and then setting $\alpha = {\left(-\dot{\imath}\right)}^{n}\,c_n$
and
$\beta = {\left(\dot{\imath}\right)}^{n}\,d_n$
in
Eq.[\ref{JKidentity}].
Next we sum over $n$  from
$n= -\infty,\ldots, \infty$
and then re-label indices so that we may identify coefficients of
$
\sum
\left[
\left(\alpha  K_{{n}+\dot{\imath}\,\nu }\,+\beta  I_{{n}+\dot{\imath}\,\nu }\,\right)\,
J_{{n}}\,
\right]
$
plus a remainder, which we require to vanish.
This generates a three term recursion relation.
We find that a general solution  to
Eq.[\ref{modeD1}] is given by

\begin{eqnarray}
\label{smodeD12}
{Y}_{m_8}(t)&=&
\sum _{n\,=\,-\infty }^{\infty }
\left\{
\,{ \dot{\imath} }^{n}\;
J_n\!\left(  e^{-t/6} \, 3 \, k\,\sqrt{2 \, B }  \right) \,
\left[
\,d_n\,I_{n+{\dot{\imath}}\nu}\left(  e^{t/6} \, 3 \, w\,\sqrt{2 \, B }  \right)\;
\right.
\right.
\nonumber \\
&+&\;
\left.
\left.
\, {\left(-1\right)}^{n}\,c_n\,\,
K_{n+{\dot{\imath}}\nu}\left(  e^{t/6} \, 3 \, w\,\sqrt{2 \, B }  \right)
\right]
\right\}
.
\end{eqnarray}

The coefficient sets $\{c_n,\,d_n\}\,,\;n= -\infty,\ldots, \infty$ are solutions of the same recurrence  relation
\begin{equation}
\label{recurrence}
\frac{1}{2}\dot{\imath}\, k\, w \, B \left(C_{n-1}+C_{n+1}\right)-
\left[ {m_{\textrm{eff}}}^2 +\frac{1}{9} \left(n+\frac{\dot{\imath}\,\nu }{2}\right)^2\right] C_n = 0,
\end{equation}
but with possibly distinct initial values, since
a general solution to the
three-term recurrence  relation Eq.[\ref{recurrence}]
possesses two arbitrary constants.

Eq.[\ref{smodeD12}] may find application in calculating
initial values in a numerical simulation,
or in the computation of the approximate cross section for the creation of
particle/anti-particle pairs of $\psi$ particles
through the annihilation of  $\varphi$ quanta, if one generalizes this
model to include an interaction of $\psi$ with the inflaton $\varphi$ of the
form
$\lambda\,\psi^*\,\psi\,\varphi$.

\newpage
\bibliographystyle{unsrt}
\bibliography{NASH_2014_CMP}
\newpage
\begin{figure}
\label{fig1}
\ifthenelse{ \equal{\USINGhyperrefNOW}{true} }{\includegraphics[width=35pc]{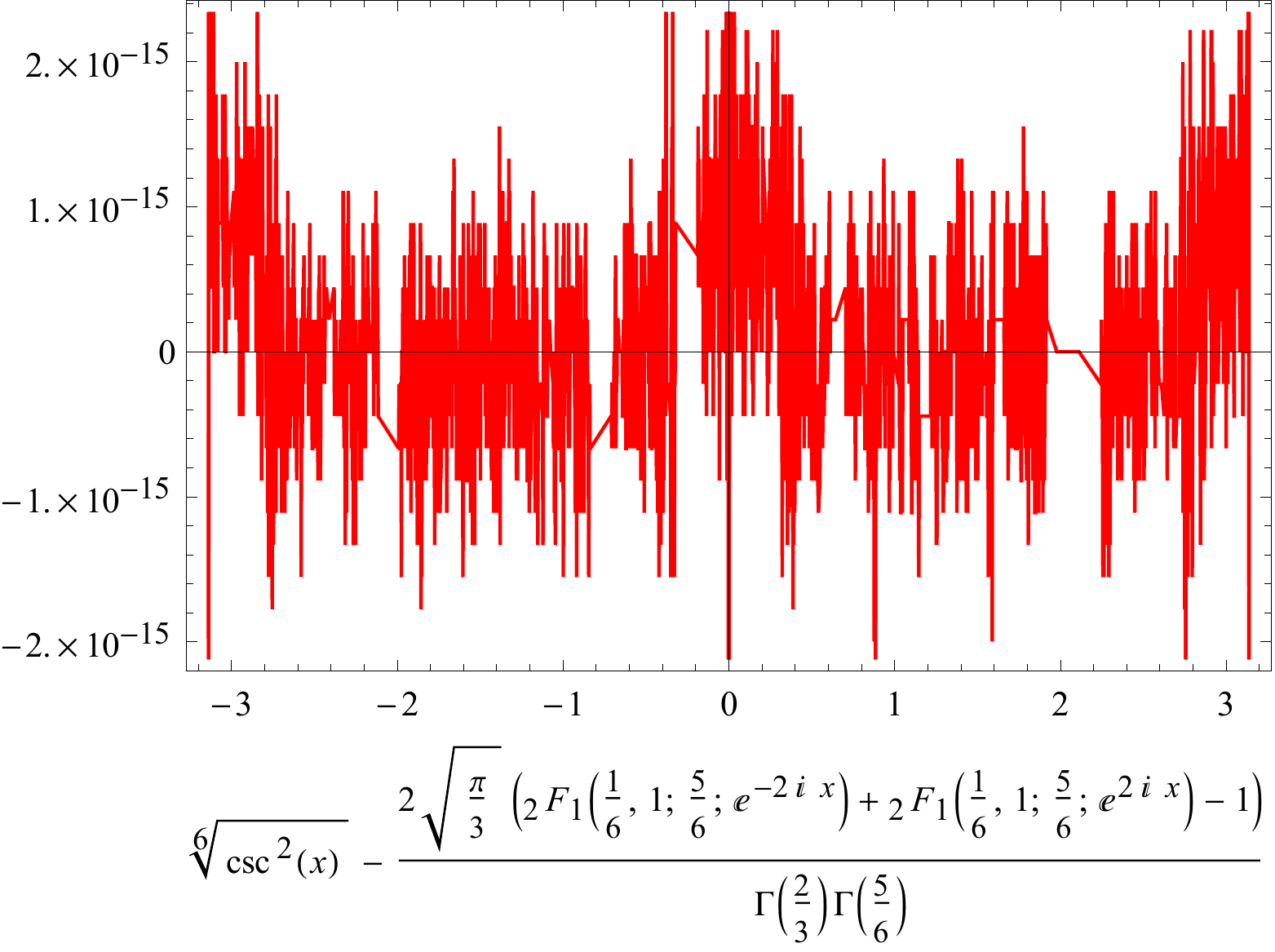}}{\includegraphics[width=35pc]{csc.eps}}
\\
\caption{Difference of $\sqrt[6]{\csc^2 (x)}$
and its realization in terms of a Fourier series that may be
evaluated in closed form using hypergeometric functions.
[Color online]}
\end{figure}
\begin{figure}
\label{fig2}
\ifthenelse{ \equal{\USINGhyperrefNOW}{true} }{\includegraphics[width=35pc]{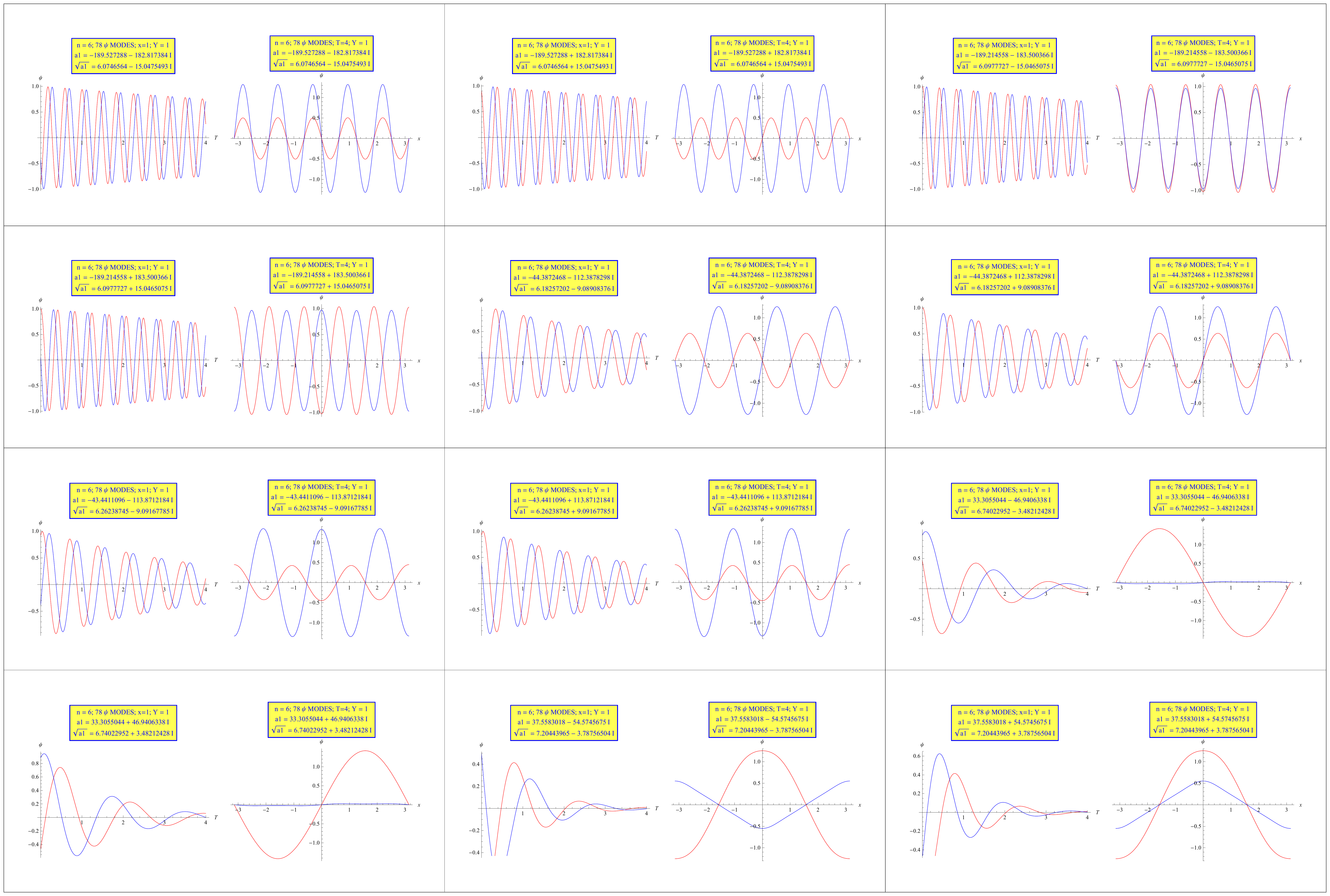}}{\includegraphics[width=35pc]{wwwg11s_read05MAY2014-n=6-reducedCut-001.eps}}
\\
\caption{Stable approximate modes FOR ODD $n_8$;
pairs of plots ($\tau$ dependence, for fixed $x$  and $x$ dependence, for fixed $\tau$).
Total number of  modes = 169 = 91 EVEN + 78 ODD n.
$n = 6, m = 0, k w \frac{\sqrt[3]{2} \Gamma \left(\frac{2}{3}\right)}{\Gamma \left(\frac{5}{6}\right)^2}=2 \sqrt{Y}=2$.
The rows and columns are delimited by distinct values of $a_1$.
[$a_1$ corresponds to odd $n_8$; $a_2$ corresponds to even $n_8$].
[Color online]}
\end{figure}
\begin{figure}
\label{fig3}
\ifthenelse{ \equal{\USINGhyperrefNOW}{true} }{\includegraphics[width=35pc]{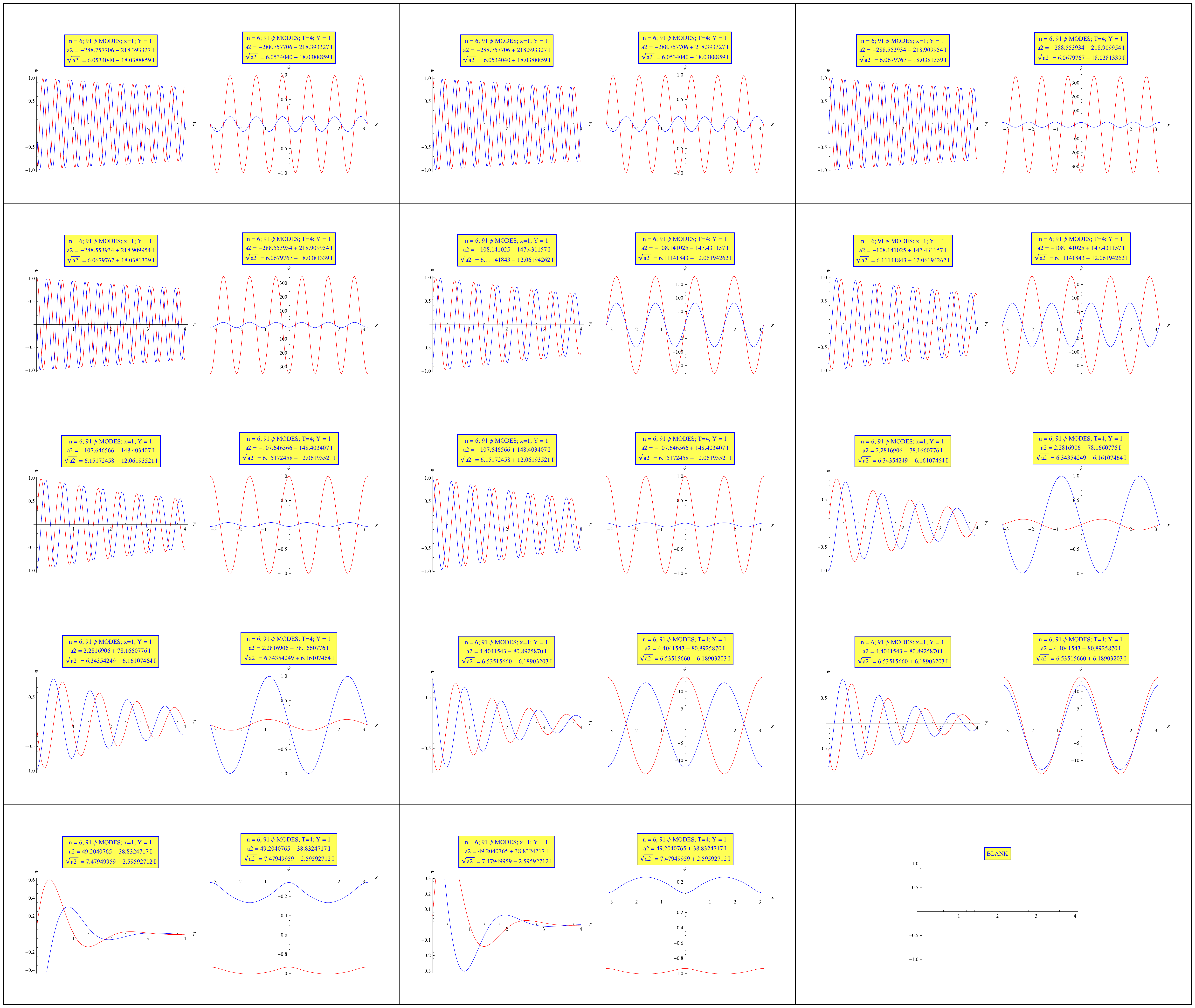}}{\includegraphics[width=35pc]{wwwg22s_read05MAY2014-n=6-reducedCut-001.eps}}
\\
\caption{Stable approximate modes FOR EVEN $n_8$;
pairs of plots ($\tau$ dependence, for fixed $x$  and $x$ dependence, for fixed $\tau$).
Total number of  modes = 169 = 91 EVEN + 78 ODD n.
$n = 6, m = 0, k w \frac{\sqrt[3]{2} \Gamma \left(\frac{2}{3}\right)}{\Gamma \left(\frac{5}{6}\right)^2}=2 \sqrt{Y}=2$.
The rows and columns are delimited by distinct values of $a_2$.
[$a_1$ corresponds to odd $n_8$; $a_2$ corresponds to even $n_8$].
[Color online]}
\end{figure}

\begin{figure}
\label{fig4}
\ifthenelse{ \equal{\USINGhyperrefNOW}{true} }{\includegraphics[width=35pc]{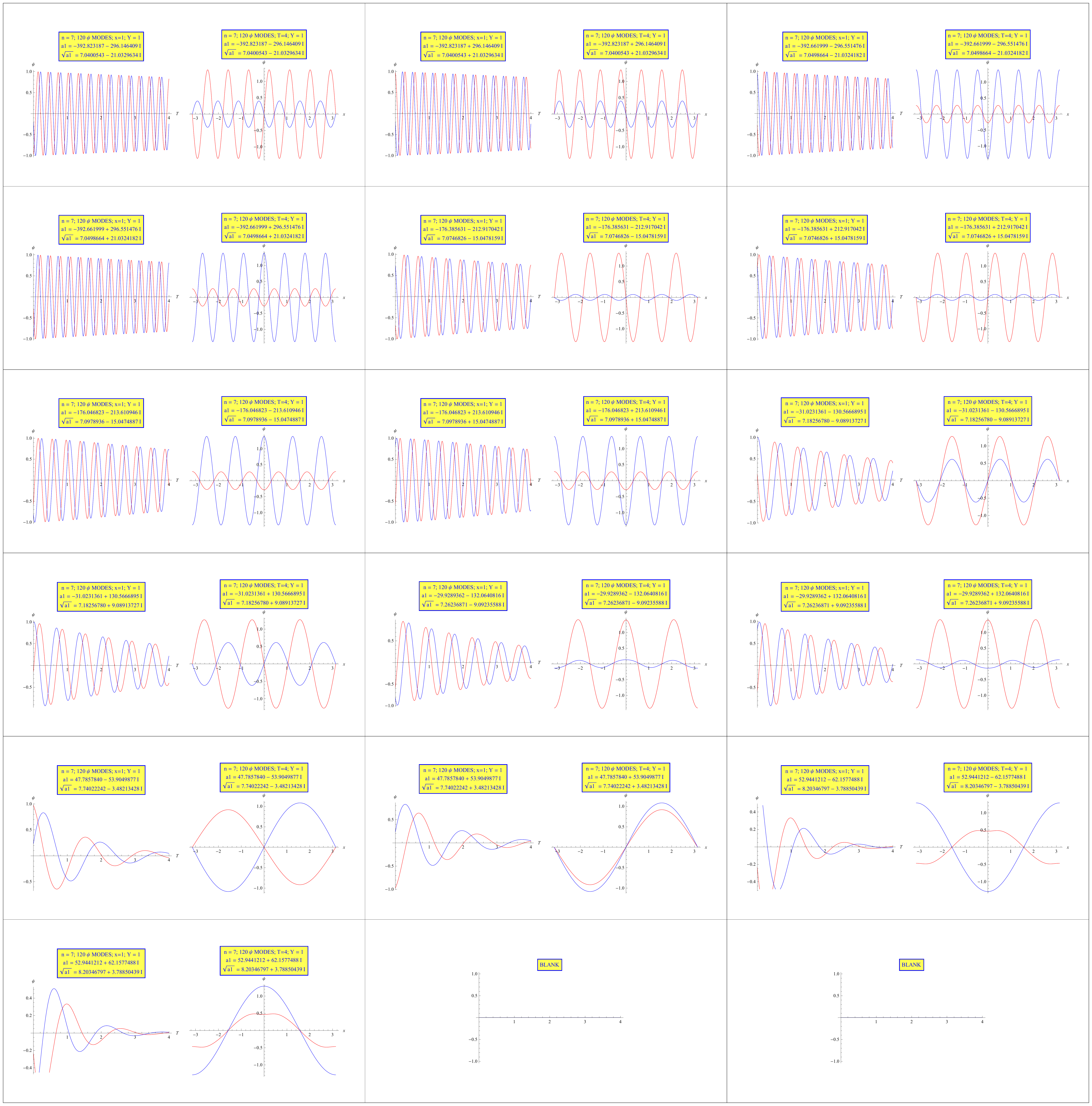}}{\includegraphics[width=35pc]{wwwg11s_05MAY2014-n=7-reducedCut-001.eps}}
\\
\caption{Stable approximate modes FOR ODD $n_8$;
pairs of plots ($\tau$ dependence, for fixed $x$  and $x$ dependence, for fixed $\tau$).
Total number of  modes = 225 = 105 EVEN + 120 ODD n.
$n = 7, m = 0, k w \frac{\sqrt[3]{2} \Gamma \left(\frac{2}{3}\right)}{\Gamma \left(\frac{5}{6}\right)^2}=2 \sqrt{Y}=2$.
The rows and columns are delimited by distinct values of $a_1$.
[$a_1$ corresponds to odd $n_8$; $a_2$ corresponds to even $n_8$].
[Color online]}
\end{figure}
\begin{figure}
\label{fig5}
\ifthenelse{ \equal{\USINGhyperrefNOW}{true} }{\includegraphics[width=35pc]{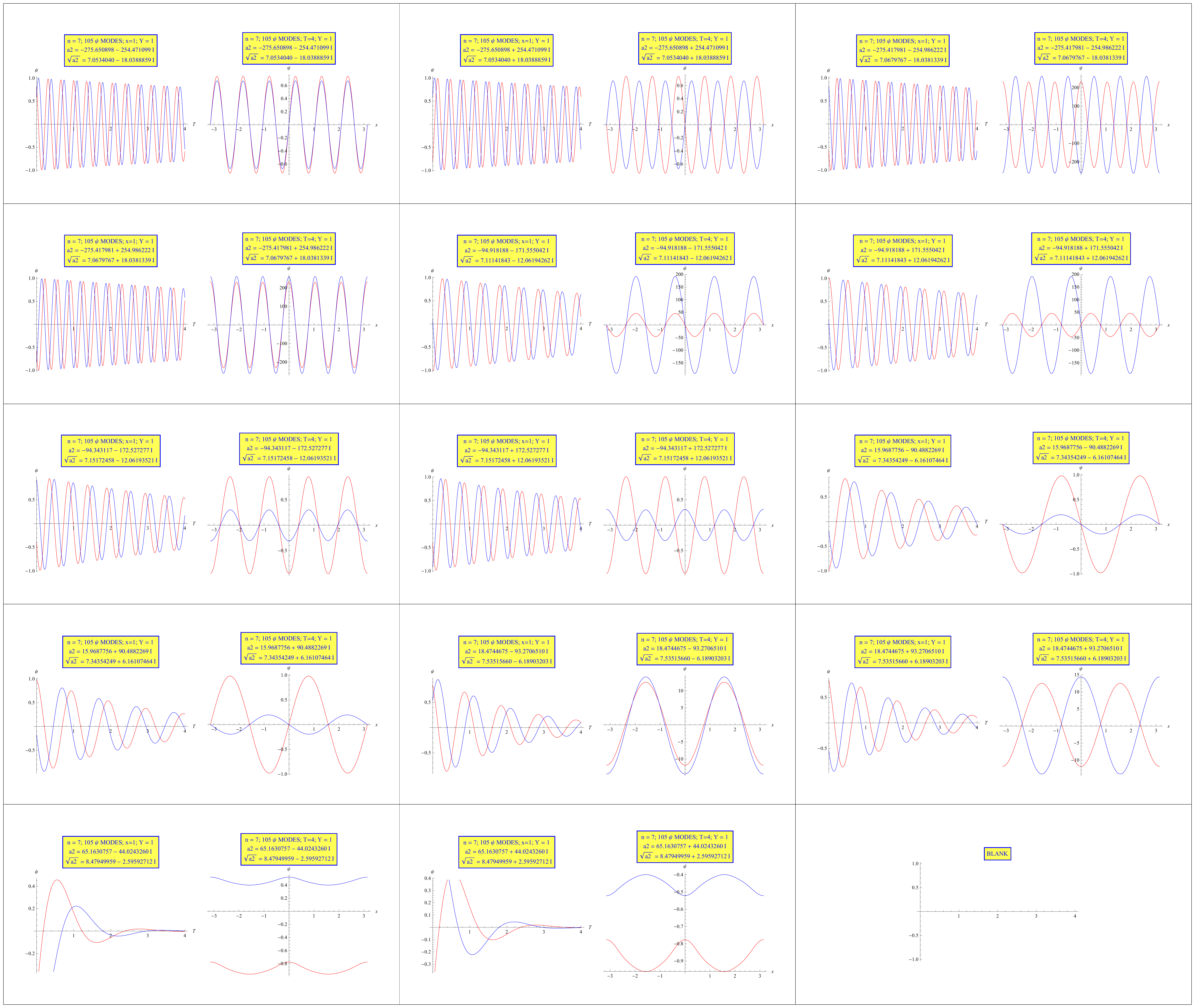}}{\includegraphics[width=35pc]{wwwg22s_05MAY2014-n=7-reducedCut-001.eps}}
\\
\caption{Stable approximate modes FOR EVEN $n_8$;
pairs of plots ($\tau$ dependence, for fixed $x$  and $x$ dependence, for fixed $\tau$).
Total number of  modes = 225 = 105 EVEN + 120 ODD n.
$n = 7, m = 0, k w \frac{\sqrt[3]{2} \Gamma \left(\frac{2}{3}\right)}{\Gamma \left(\frac{5}{6}\right)^2}=2 \sqrt{Y}=2$.
The rows and columns are delimited by distinct values of $a_2$.
[$a_1$ corresponds to odd $n_8$; $a_2$ corresponds to even $n_8$].
[Color online]}
\end{figure}

\begin{figure}
\label{fig6}
\ifthenelse{ \equal{\USINGhyperrefNOW}{true} }{\includegraphics[width=35pc]{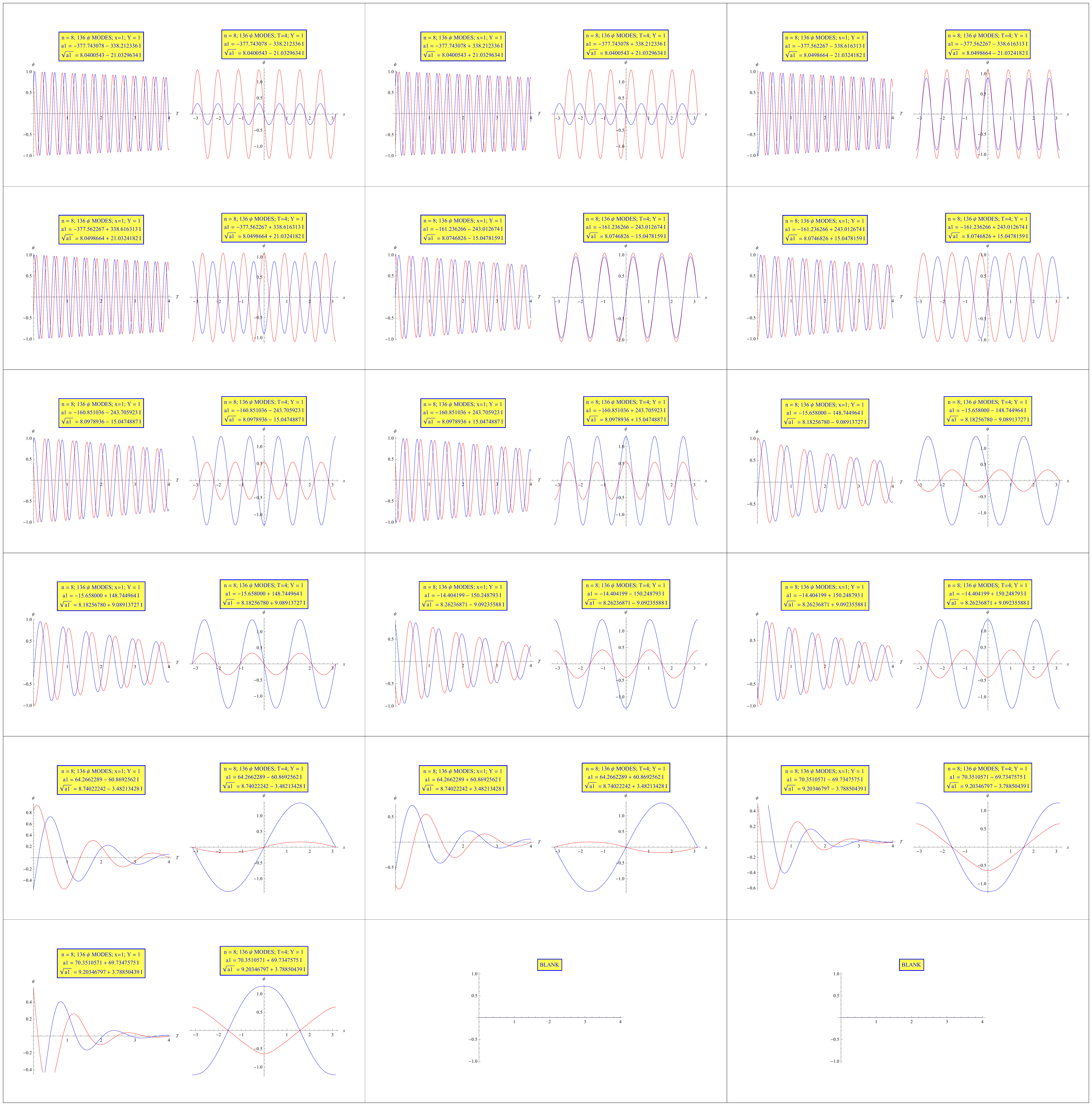}}{\includegraphics[width=35pc]{wwwg11s_read10MAY2014-n=8-01.eps}}
\\
\caption{Stable approximate modes FOR ODD $n_8$;
pairs of plots ($\tau$ dependence, for fixed $x$  and $x$ dependence, for fixed $\tau$).
Total number of  modes = 289 = 153 EVEN + 136 ODD n.
$n = 8, m = 0, k w \frac{\sqrt[3]{2} \Gamma \left(\frac{2}{3}\right)}{\Gamma \left(\frac{5}{6}\right)^2}=2 \sqrt{Y}=2$.
The rows and columns are delimited by distinct values of $a_1$.
[$a_1$ corresponds to odd $n_8$; $a_2$ corresponds to even $n_8$].
[Color online]}
\end{figure}
\begin{figure}
\label{fig7}
\ifthenelse{ \equal{\USINGhyperrefNOW}{true} }{\includegraphics[width=35pc]{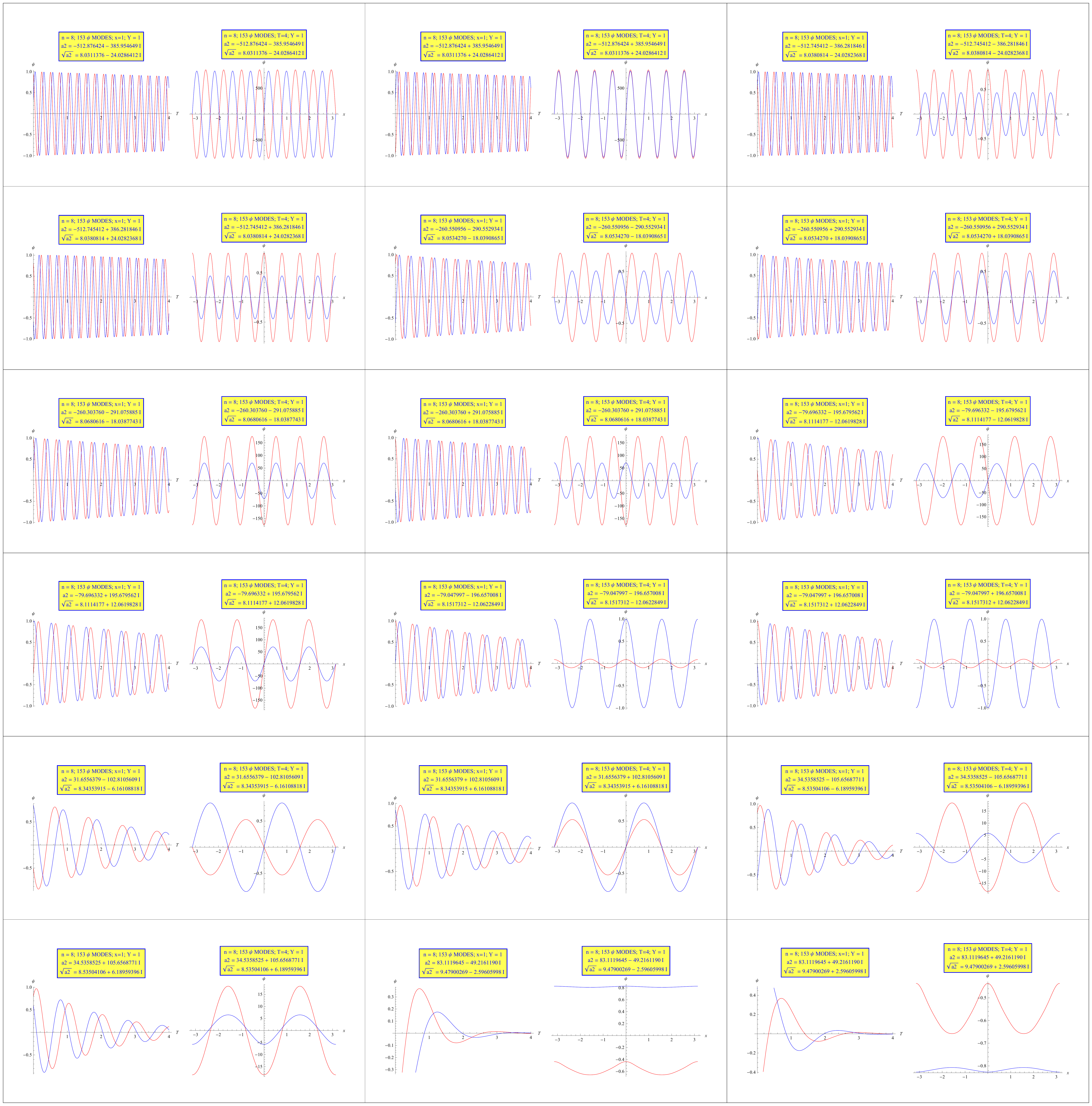}}{\includegraphics[width=35pc]{wwwg22s_read10MAY2014-n=8-01.eps}}
\\
\caption{Stable approximate modes FOR EVEN $n_8$;
pairs of plots ($\tau$ dependence, for fixed $x$  and $x$ dependence, for fixed $\tau$).
Total number of  modes = 289 = 153 EVEN + 136 ODD n.
$n = 8, m = 0, k w \frac{\sqrt[3]{2} \Gamma \left(\frac{2}{3}\right)}{\Gamma \left(\frac{5}{6}\right)^2}=2 \sqrt{Y}=2$.
The rows and columns are delimited by distinct values of $a_2$.
[$a_1$ corresponds to odd $n_8$; $a_2$ corresponds to even $n_8$].
[Color online]}
\end{figure}

\end{document}